\documentclass{aastex62}
\usepackage{lipsum} 
\usepackage{footnote}
\usepackage{fancyhdr}
\shorttitle{AMMONIA MAP OF L1495-B218: II CCS, HC$_7$N, and STAR FORMATION}
\shortauthors{Seo et al.}

\begin{document}

\title{AN AMMONIA SPECTRAL MAP OF THE L1495-B218 FILAMENTS IN THE TAURUS MOLECULAR CLOUD: II. CCS \& HC$_7$N CHEMISTRY AND THREE MODES OF STAR FORMATION IN THE FILAMENTS}


\author{Young Min Seo}
\affiliation{Jet Propulsion Laboratory, California Institute of Technology, 4800 Oak Grove Drive, Pasadena, CA, 91109, USA}

\author{Liton Majumdar}
\affiliation{Jet Propulsion Laboratory, California Institute of Technology, 4800 Oak Grove Drive, Pasadena, CA, 91109, USA}

\author{Paul F. Goldsmith}
\affiliation{Jet Propulsion Laboratory, California Institute of Technology, 4800 Oak Grove Drive, Pasadena, CA, 91109, USA}

\author{Yancy L. Shirley}
\affiliation{Department of Astronomy \& Steward Observatory, University of Arizona, 933 N. Cherry Ave., Tucson, AZ 85721, USA}

\author{Karen Willacy}
\affiliation{Jet Propulsion Laboratory, California Institute of Technology, 4800 Oak Grove Drive, Pasadena, CA, 91109, USA}

\author{Derek Ward-Thompson}
\affiliation{Jeremiah Horrocks Institute, University of Central Lancashire, Preston PR1 2HE, UK}

\author{Rachel Friesen}
\affiliation{National Radio Astronomy Observatory, 520 Edgemont Rd., Charlottesville, VA 22903, USA}

\author{David Frayer}
\affiliation{Green Bank Observatory, 155 Observatory Road, Green Bank, WV 24944, USA}

\author{Sarah E. Church}
\affiliation{Kavli Institute for Particle Astrophysics and Cosmology; Physics Department, Stanford University, Stanford, CA 94305, USA}

\author{Dongwoo Chung}
\affiliation{Kavli Institute for Particle Astrophysics and Cosmology; Physics Department, Stanford University, Stanford, CA 94305, USA}

\author{Kieran Cleary}
\affiliation{California Institute of Technology, Pasadena, CA, 91125, USA}

\author{Nichol Cunningham}
\affiliation{Green Bank Observatory, 155 Observatory Road, Green Bank, WV 24944, USA}
\affiliation{Institut de Radioastronomie Millimetrique (IRAM), 300 rue de la Piscine, 38406 Saint Martin d'H$\grave{e}$res, France}

\author{Kiruthika Devaraj}
\affiliation{Kavli Institute for Particle Astrophysics and Cosmology; Physics Department, Stanford University, Stanford, CA 94305, USA}

\author{Dennis Egan}
\affiliation{Green Bank Observatory, 155 Observatory Road, Green Bank, WV 24944, USA}

\author{Todd Gaier}
\affiliation{Jet Propulsion Laboratory, California Institute of Technology, 4800 Oak Grove Drive, Pasadena, CA, 91109, USA}

\author{Rohit Gawande}
\affiliation{Jet Propulsion Laboratory, California Institute of Technology, 4800 Oak Grove Drive, Pasadena, CA, 91109, USA}

\author{Joshua O. Gundersen}
\affiliation{Department of Physics, University of Miami, 1320 Campo Sano Drive, Coral Gables, FL 33146, USA}

\author{Andrew I. Harris}
\affiliation{Department of Astronomy, University of Maryland, College Park, MD, 20742, USA}

\author{Pekka Kangaslahti}
\affiliation{Jet Propulsion Laboratory, California Institute of Technology, 4800 Oak Grove Drive, Pasadena, CA, 91109, USA}

\author{Anthony C.S. Readhead}
\affiliation{California Institute of Technology, Pasadena, CA, 91125, USA}

\author{Lorene Samoska}
\affiliation{Jet Propulsion Laboratory, California Institute of Technology, 4800 Oak Grove Drive, Pasadena, CA, 91109, USA}

\author{Matthew Sieth}
\affiliation{Kavli Institute for Particle Astrophysics and Cosmology; Physics Department, Stanford University, Stanford, CA 94305, USA}

\author{Michael Stennes}
\affiliation{Green Bank Observatory, 155 Observatory Road, Green Bank, WV 24944, USA}

\author{Patricia Voll}
\affiliation{Kavli Institute for Particle Astrophysics and Cosmology; Physics Department, Stanford University, Stanford, CA 94305, USA}

\author{Steve White}
\affiliation{Green Bank Observatory, 155 Observatory Road, Green Bank, WV 24944, USA}

\begin{abstract}

We present deep CCS and HC$_7$N observations of the L1495-B218 filaments in the Taurus molecular cloud obtained using the K-band focal plane array on the 100m Green Bank Telescope. We observed the L1495-B218 filaments in CCS $J_N$ = 2$_1$$-$1$_0$ and HC$_7$N $J$ = 21$-$20 with a spectral resolution of 0.038 km s$^{-1}$ and an angular resolution of 31$''$. We observed strong CCS emission in both evolved and young regions and weak emission in two evolved regions. HC$_7$N emission is observed only in L1495A-N and L1521D. We find that CCS and HC$_7$N intensity peaks do not coincide with NH$_3$ or dust continuum intensity peaks. We also find that the fractional abundance of CCS does not show a clear correlation with the dynamical evolutionary stage of dense cores. Our findings and chemical modeling indicate that the fractional abundances of CCS and HC$_7$N are sensitive to the initial gas-phase C/O ratio, and they are good tracers of young condensed gas only when the initial C/O is close to solar value. Kinematic analysis using multiple lines including NH$_3$, HC$_7$N, CCS, CO, HCN, \& HCO$^+$ suggests that there may be three different star formation modes in the L1495-B218 filaments. At the hub of the filaments, L1495A/B7N has formed a stellar cluster with large-scale inward flows (fast mode), while L1521D, a core embedded in a filament, is slowly contracting due to its self-gravity (slow mode). There is also one isolated core that appears to be marginally stable and may undergo quasi-static evolution (isolated mode).

\end{abstract}

\keywords{ISM: clouds --- ISM: molecules $-$- radio lines: ISM --- stars: formation --- surveys}

\section{Introduction} \label{sec:intro}

Star formation is one of the fundamental processes in the Universe because it drives the evolution of galaxies and sets up the initial conditions for planet formation and the emergence of life. Star formation begins with gravitational contraction of a molecular cloud forming progressively smaller and denser structures such as molecular filaments and dense cores. If a dense core becomes gravitationally unstable, it forms a protostar or a group of protostars. While this general framework is established, the details of the star formation processes are still poorly understood. Particularly, how a molecular cloud evolves to filaments and dense cores and how dense cores within filaments interact with their surroundings and evolve to form stars are key questions. Answering these questions has been challenging due to difficulties in observations.

The main difficulty originates from making complete and consistent observations toward structures over a wide range of physical properties. Molecular clouds are typically a few tens of parsecs in size with an average density of 100 cm$^{-3}$. Molecular cloud cores are less than 0.5 parsecs in size with central densities ranging from 10$^4$ cm$^{-3}$ to $>$10$^6$ cm$^{-3}$. The first hydrostatic cores (FHSC) are expected to be a few AU in size with a density of 10$^{10}$ cm$^{-3}$ and a temperature of 2000 K. A sub-mm/mm-wavelength continuum observation may trace dust structures over a wide range of column densities from a molecular cloud to an FHSC but it does not yield any velocity information. On the other hand, molecular lines give velocity information and facilitate identification of complex three dimensional structures. However, the range of physical properties traced by a single molecular line is firmly limited by the chemical and excitation characteristics of the molecule. Thus, it is crucial to know which physical and chemical condition of complex structures can be correctly traced by a specific molecular line and how to interpret the kinematic information.

In this study, we make observations in CCS $J_N$ = 2$_1$ $-$ 1$_0$ toward the L1495-B218 filaments in Taurus. We investigate what CCS $J_N$ = 2$_1$ $-$ 1$_0$ is tracing in the region and discuss various modes of star formation in the L1495-B218 filaments using complementary spectral observations of HC$_7$N, HCN, HCO$^+$ and CO isotopologues.

CCS is a carbon-chain molecule which is frequently found in dense cores in nearby, low-mass star-forming clouds \citep[e.g.,][]{suzuki92, wolkovitch97, rathborne08, roy11, marka12, tatematsu14a}. Yet, what physical and chemical properties of star-forming clouds can be probed by CCS and how CCS traces the chemical evolution during star formation are still debated. \citet{suzuki92} made the first CCS survey toward 49 dense cores in Taurus and Ophiuchus as well as a chemical evolution model of CCS with respect to the dense gas tracers NH$_3$ and N$_2$H$^+$. They found that carbon chain molecules including CCS and cyanopolyynes (HC$_X$N; X=3,5,7,$\ldots$) are less abundant in star forming regions where NH$_3$ emission is strong. Their chemical model also predicts that the abundance ratio of [CCS]/[NH$_3]$ has an anti-correlation with time due to CCS depletion \citep{suzuki92, aikawa01}. Some surveys toward Bok globules and dense cores supported the chemical model of \citet{suzuki92} by showing a trend of low [CCS]/[NH$_3$] ratios in protostallar cores but high [CCS]/[NH$_3$] ratios in starless cores \citep{scappini96, gregorio06, foster09, tatematsu14b}.

Several observations have results that are not consistent with the chemical model of \citet{suzuki92}. \citet{benson98} found no clear trend of [CCS]/[N$_2$H$^+$] as a function of the evolutionary stage of the 20 dense cores with both CCS and N$_2$H$^+$ emission in their survey of 60 dense cores. \citet{marka12} also showed that the isolated Bok globules in their survey do not have a clear anti-correlation of the N(CCS)/N(NH$_3$) ratio with evolutionary state moving from starless cores to protostellar cores. Observations toward infrared dark clouds (IRDCs) in NH$_3$ and CCS using the GBT and the VLA do not show any evidence of a trend in N(CCS)/N(NH$_3$) from starless to protostellar clumps but rather have highly scattered values \citep{dirienzo15}. Furthermore, a quantitative analysis of the fractional abundances of CCS, NH$_3$, HCO$^+$, CO, etc. from spatially resolved core observations shows considerably inconsistencies with Suzuki et al.'s prediction and suggests that either the model of chemical processes in a dense core needs to be revised \citep{lee03} or that a dynamically evolving dense core (e.g., a core with infall motion) has a considerably different chemical evolutionary path than that of a static core \citep{shematovich03}.

We use the L1495-B218 filaments in Taurus to study the physical and chemical properties of CCS and HC$_7$N and to investigate the star formation processes in the Taurus molecular cloud using complementary high-resolution observations including HCN, HCO$^+$ and CO isotopologues. The L1495-B218 filaments in Taurus, which consist of multiple velocity-coherent filaments \citep{hacar13}, are good test beds because their structures and evolutionary stages are well observed and determined from large to dense core scale. \citet{goldsmith08} observed large-scale structures in the Taurus molecular cloud including the L1495-B218 filaments in $^{12}$CO and $^{13}$CO $J$ = 1$-$0. Their observations also showed that there is a converging or a shear flow toward the filaments \citep{narayanan08, palmeirim13}, which is likely augmenting the mass of the filaments. Protostars are surveyed and identified in IR using, e.g., the Spitzer Space Telescope \citep{rebull10, rebull11}. L1495, B213, and B218 have at least one Class I or Flat spectrum protostar while B10, B211, and B216 have no protostar or only one Class II or III protostar, which suggests that L1495/B7, B213, and B218 are relatively more evolved than B10, B211, and B216 in terms of star formation \citep{seo15}. Having various evolutionary stages in Taurus enables us to trace the evolutionary processes of low-mass star formation.

Smaller structures in the L1495-B218 regions have been identified using gas velocity information obtained from observations of C$^{18}$O and NH$_3$ \citep{hacar13, seo15}. Thirty-five velocity coherent filaments were identified across the L1495-B218 regions and contain a mixture of gravitationally stable and unstable filaments with line density ranging from 3 M$_\odot$ pc$^{-1}$ to 71 M$_\odot$ pc$^{-1}$ \citep{hacar13} (the critical line density is 16 M$_\odot$ pc$^{-1}$ at 10 K, \citealt{ostriker64}). Thirty-nine NH$_3$ cores within the filaments are identified and studied in NH$_3$ \citep[hereafter Paper I]{seo15} which have good spatial coincidences with bright dust cores. The dense cores show various evolutionary stages, including 35 starless cores and four protostellar cores \citep{rebull10, rebull11}. The L1495-B218 filaments and dense cores share a similar surrounding environment, making a consistent survey of the physical and chemical properties of CCS and star formation process possible.

We mapped the three-degree long L1495-B218 filaments in CCS $J_N$ = 2$_1$ $-$ 1$_0$ and HC$_7$N $J$ = 21$-$20 using the Robert C. Byrd Green Bank Telescope (GBT). We analyze the kinematics of CCS-bright regions and discuss what CCS emission tells us about dense core evolution and star formation. We also analyze the cyanopolyyne HC$_7$N emission that was in the bandpass of previously-published NH$_3$ observations in \citet{seo15}. converging-motion tracers, HCN and HCO$^+$ $J$ = 1$-$0, were observed toward 26 dense cores including three dense cores with strong CCS emission using the 12m radio telescope of the Arizona Radio Observatory and the Robert C. Byrd Green Bank Telescope (GBT). The layout of this paper is as follows: \S2 briefly introduces our mapping of the L1495-B218 filaments in CCS and the data reduction procedures. In \S3 we present the basic physical properties of the L1495-B218 filaments seen in CCS and kinematics of three CCS-bright regions. In \S4 we discuss what CCS traces in a molecular cloud and the physical states of CCS-bright regions by analyzing CCS with $^{12}$CO, $^{13}$CO, C$^{18}$O, NH$_3$, HCN, and HCO$^+$. Finally, we summarize our results in \S5.

\section{OBSERVATION \& DATA REDUCTION }\label{sec:obs}

\subsection{Mapping in CCS $J_N$ = 2$_1$$-$1$_0$ and HC$_7$N $J$ = 21$-$20}\label{sec:obs:CCS}

\begin{figure*}[htb!]
\includegraphics[angle=0, scale=0.88]{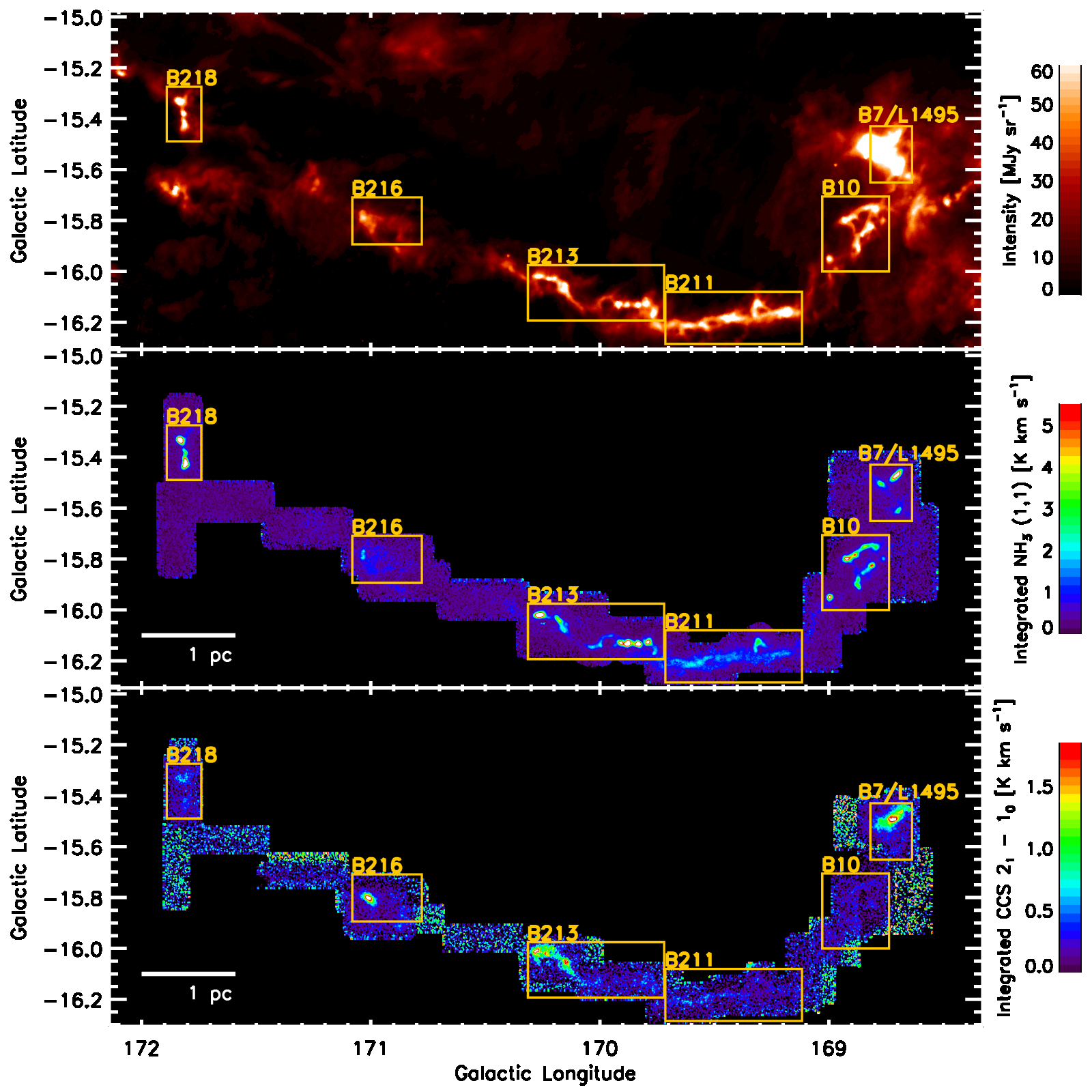}
\caption{Top: 500 $\mu m$ dust continuum emission seen by the \textit{SPIRE} instrument on the \textit{Herschel Space Observatory} \citep{palmeirim13}. Middle : map of integrated intensity of NH$_3$ (1,1) \citep{seo15}. Bottom: map of integrated intensity of CCS $J_N$ = 2$_1$$-$1$_0$. \label{totalmap}}
\end{figure*}
\begin{figure*}[htb!]
\includegraphics[angle=0,scale=0.87]{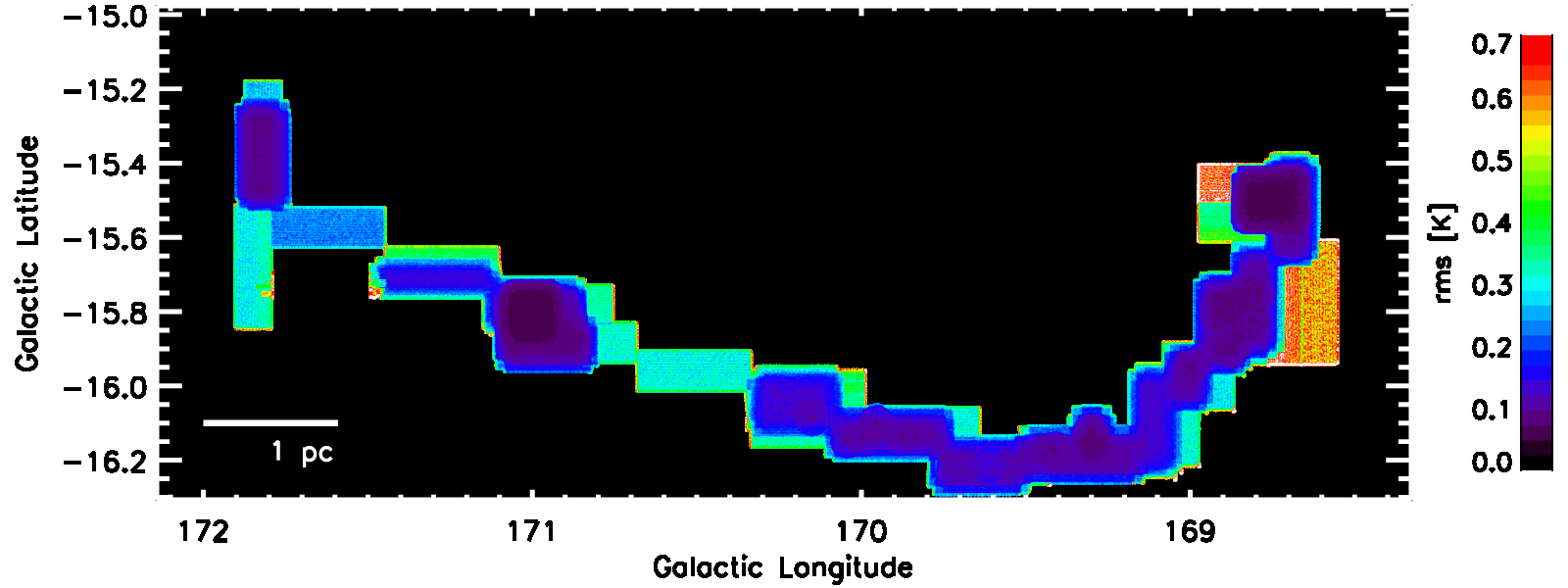}
\caption{$rms$ of the CCS map in $T_{\rm mb}$ at a spectral resolution of 6.1 kHz (corresponding to 0.082 km s$^{-1}$), obtained by smoothing two spectral channels. \label{rms}}
\end{figure*}

We observed the L1495-B218 filaments using the GBT in CCS $J_N$ = 2$_1$$-$1$_0$ (GBT12A-295 \& GBT13A-126) in spring of 2012 and 2013. In the first year, we mapped the filaments in NH$_3$ (1,1) and (2,2) and HC$_7$N $J$ = 21$-$20 using 7 beams of the K-band focal plane array (KFPA) and in CCS $J_N$ = 2$_1$$-$1$_0$ (22.344033 GHz; \citealt{yamamoto90}) using one beam of the KFPA (7+1 mode) from April 2012 to Feb 2013 (see Paper I). In the second year of observations, we mapped the L1495-B218 filaments in CCS 2$_1$$-$1$_0$ using seven beams. We mapped each region using position switching. Three OFF positions were selected having extinction small enough not to have CCS 2$_1$$-$1$_0$ emission. Pointing of the telescope was updated by observing the quasar 0403+2600 every two hours. To calibrate the gain of seven KFPA beams, we observed either Venus or Jupiter in every observing shift and used the same calibration procedure described in \citet{seo15} (hereafter Paper I) to place spectra on the T$_{\rm mb}$ scale. We also observed the peak dust continuum position of the L1489PPC starless core ($\alpha$ = 04:04:47.6, $\delta$ = +26:19:17.9, $J$2000.0; \citealt{young04,ford11}) in every observing shift for a secondary check on the consistency of the absolute flux calibration. The FWHM beam size of the GBT at 22 GHz is 31$''$.

The CCS data have been reduced in the same way as described in Paper I. Figure \ref{totalmap} shows the integrated intensity map of CCS $J_N$ = 2$_1$$-$1$_0$ and Figure \ref{rms} shows the $rms$ of the CCS map. We deeply mapped the regions with the NH$_3$ (1,1) emission above 300 mK and the region with the 500 $\mu$m continuum emission above 20 MJy sr$^{-1}$. We used all seven beams of the KFPA and achieved an $rms$ noise of $\leq$200 mK. The median noise level is 154 mK with the lowest noise level of 26 mK in L1495.

The HC$_7$N data have been reduced in the same way as described in Paper I. The $rms$ of the HC$_7$N map is the same as the $rms$ map of NH$_3$ shown in Paper I since HC$_7$N $J$ = 21$-$20 is in the same bandpass as NH$_3$ (1,1) and (2,2).

\subsection{HCN \& HCO$^+$ 1$-$0 Observations}\label{sec:obs:HCN}

We observed the 89 GHz transitions of HCN and HCO$^+$ to probe the internal dynamics of dense cores. We surveyed 26 dense cores including L1495A-N, L1521B, and L1521D in HCN $J$ = 1$-$0 (88.631847 GHz) and HCO$^+$ $J$ = 1$-$0 (89.1885 GHz) in the Winters of 2013, 2016 and 2017. We used the 12m Arizona Radio Observatory (ARO) and the Argus receiver on the 100m GBT \citep{devaraj2015} (GBT16B-114). Coordinates of L1495A-N, L1521B, and L1521D are ($\alpha$, $\delta$) = 4:18:33.0, +28:27:58, ($\alpha$, $\delta$) = 4:24:20.6, +26:36:02, ($\alpha$, $\delta$) = 4:21:21.7, +26:59:31, respectively, in $J$2000.0. The details of the HCN and HCO$^+$ survey will be reported in a separate paper.

The ARO has a single beam with FWHM beam size of 66$''$ at 89 GHz. For an accurate calibration of the 12m ARO data, we observed Jupiter every 1.5 hours to adjust focus and pointing and to calibrate the spectra onto the T$_{\rm mb}$ scale. The system temperatures were around 180 K in T$_{\rm A}$. The main beam efficiency was typically $\simeq$0.6. A linear baseline was removed by the position switching. The total integration time for each target was 30 minutes with a typical $rms$ of $<$120 mK in units of T$_{\rm mb}$ unit at a spectral resolution of 0.041 km s$^{-1}$. We reduced the data using the CLASS software \citep{pety05}.

Argus is a 16-pixel W-band array receiver with a 4x4 configuration on the GBT \citep{sieth14}. We used Beam 10 as the pointing beam on the three cores and the other beams to sparsely map the outskirts of the cores. The FWHM beam size is 9$''$ at 88 GHz. For calibration, we observed the blazar J0510+1800, which is an ALMA calibration source, to obtain the focus and pointing solutions and to calibrate the spectra in units of main beam temperature. The system temperatures were $\simeq$150 K in T$^*_{\rm A}$. The main beam efficiencies were typically 0.4 for all beams. A continuum baseline was removed by the frequency switching observation with a total shift of 7 kHz ($\pm$3.5 kHz shift). The total integration time for each target was 10 minutes with resulting a $rms$ generally less than 150 mK in T$_{\rm MB}$ at a spectral resolution of 0.041 km s$^{-1}$. We reduced the data using the GBTIDL package \citep{marganian06} provided by the Green Bank Observatory.

\section{RESULTS}

\subsection{Integrated Intensities of CCS $J_N$ = 2$_1$$-$1$_0$ and HC$_7$N $J$ = 21$-$20}

\begin{figure*}[tb]
\includegraphics[angle=0,scale=0.6]{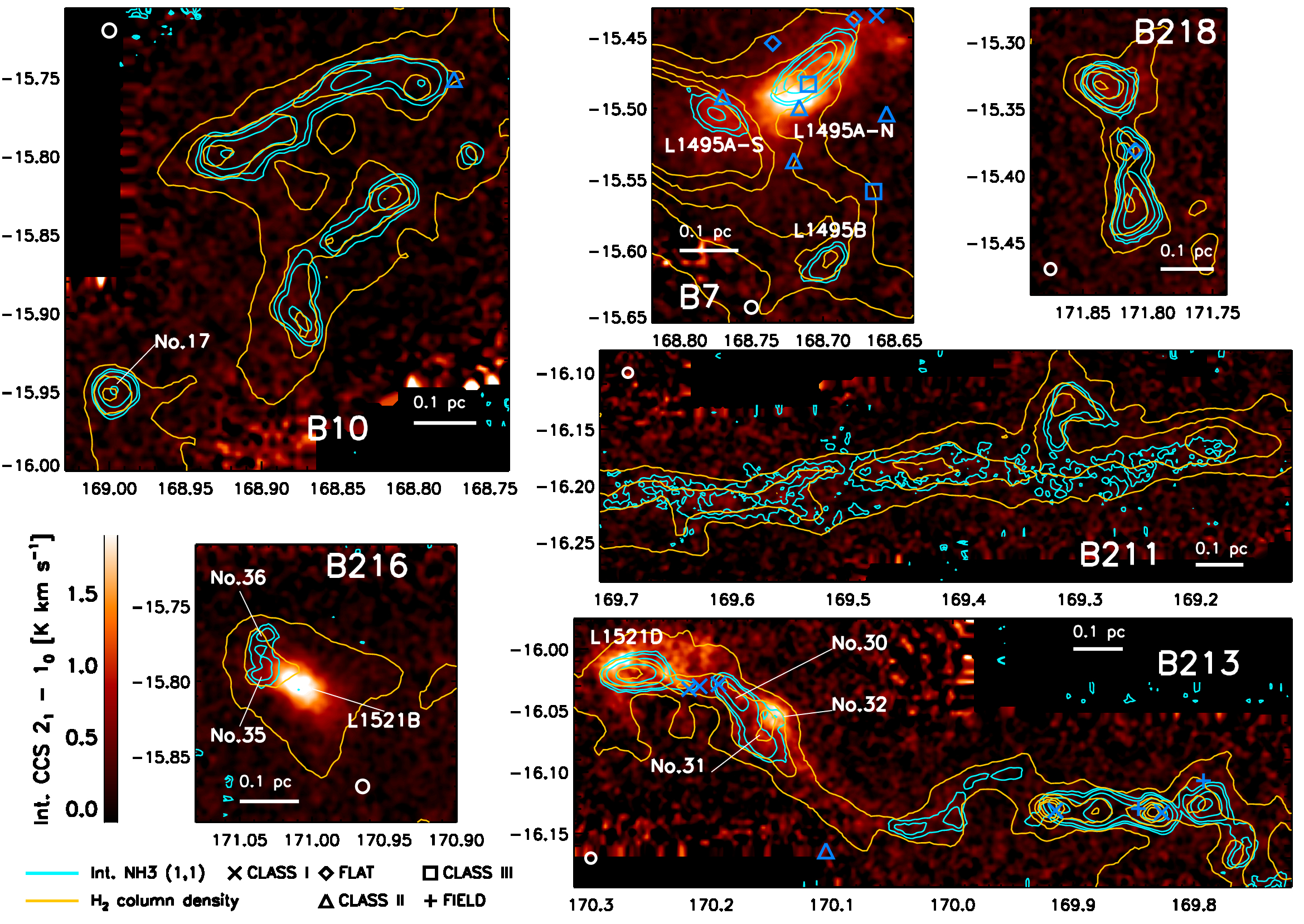}
\caption{{Images of integrated CCS 2$_1$ $-$ 1$_0$ (color), integrated NH$_3$ (1,1) (cyan contours), H$_2$ column density (orange contours), and positions of known protostars. The X-axis is Galactic longitude and Y-axis is Galactic latitude. NH$_3$ contours are at integrated intensities of 1, 1.5, 3, 6, and 12 K km s$^{-1}$. H$_2$ column density contours are at 5 $\times$ 10$^{21}$, 1 $\times$ 10$^{22}$, 1.5 $\times$ 10$^{22}$, 2 $\times$ 10$^{22}$, and 2.5 $\times$ 10$^{22}$ cm$^{-3}$. The names of previously-studied cores discussed in this study are indicated in the images. For the dense cores discussed in this study but not having classical names, we used the IDs of NH$_3$ cores in \citet{seo15}. The white circle in each image shows the beam size.}}
\label{zoomin1}
\end{figure*}

We present the integrated intensity of CCS $J_N$ = 2$_1$$-$1$_0$ emission (hereafter CCS 2$_1$$-$1$_0$) in Figures \ref{totalmap} and \ref{zoomin1}. The CCS 2$_1$$-$1$_0$ emission is bright ($>$1 K km s$^{-1}$) mainly in three regions: north of L1495/B7 (L1495A/B7N), east of B213 (hereafter B213E), and B216. There is also weak emission in B211 and B218. No significant CCS 2$_1$$-$1$_0$ emission is observed in B10 and west of B213 (hereafter B213W) at our noise level (we detect only a hint of the emission at the 2$\sigma$ level). The lowest intensity of 500 $\mu m$ dust continuum emission where the CCS 2$_1$$-$1$_0$ intensity is above 3$\sigma$ (210 mK) is 15 MJy sr$^{-1}$, while it is 25 MJy sr$^{-1}$ for NH$_3$ (1,1); however, not all structures at 15 MJy sr$^{-1}$ have CCS 2$_1$$-$1$_0$ emission. In B10, we do not have a solid detection of  CCS 2$_1$$-$1$_0$ emission at our $rms$ level of 70 mK even though the dust continuum intensity spans the range from $<$15 MJy sr$^{-1}$ to 250 MJy sr$^{-1}$. This confirms that CCS 2$_1$$-$1$_0$ emission traces lower density structures compared to NH$_3$ (1,1) but not every low density structure. For example, CCS 2$_1$$-$1$_0$ emission is bright in both more-evolved (L1495A/B7N \& B213E) and less-evolved (B216) regions, while it is weak or not detected in a couple of less evolved regions (B10 and B211) \citep[see][, for an explanation of the evolutionary stages of regions]{seo15}.

We compare our CCS 2$_1$$-$1$_0$ map to an NH$_3$ (1,1) map \citep{seo15} and a dust continuum map at 500 ${\rm \mu m}$ \citep{palmeirim13} in Figure \ref{zoomin1}. In general, the peak intensities of CCS 2$_1$$-$1$_0$ emission are displaced from the intensity peaks of NH$_3$ and dust continuum while the intensity peaks of the NH$_3$ and the 500 ${\rm \mu m}$ dust continuum are spatially well correlated.

The peak intensities of CCS 2$_1$$-$1$_0$ emission in L1521D and L1495A-N are at the outskirts of the NH$_3$ cores, $\sim$0.1 pc from the NH$_3$ (1,1) intensity peaks (2.5$'$ at the distance to Taurus, 140 pc; \citealt{loinard07, torres09}), and show an arc-shaped morphology which partially surrounds the NH$_3$ cores. CCS 2$_1$$-$1$_0$ emission is also bright at a clump located at 0.35 pc southwest of L1521D. The clump contains NH$_3$ dense cores No.30, No.31, and No.32 (see Figure 9 in Paper I). Among the three cores, CCS 2$_1$$-$1$_0$ emission is brightest at the NH$_3$ core No.32, and the position of intensity peak agrees relatively well (within 0.05 pc or 74$''$ which is a rare occurrence in this mapping survey) with the dust and NH$_3$ peaks. In B216, which is thought to be a less-evolved region, the CCS 2$_1$$-$1$_0$ peak is located about 0.1 pc west of the NH$_3$ (1,1) and dust continuum intensity peaks. Notably, the CCS 2$_1$$-$1$_0$ intensity peaks are spatially displaced from the NH$_3$ cores identified in Paper I. This suggests that CCS 2$_1$$-$1$_0$ emission does not trace the dense core in this region.

\begin{figure*}[tb]
\includegraphics[angle=0,scale=0.88]{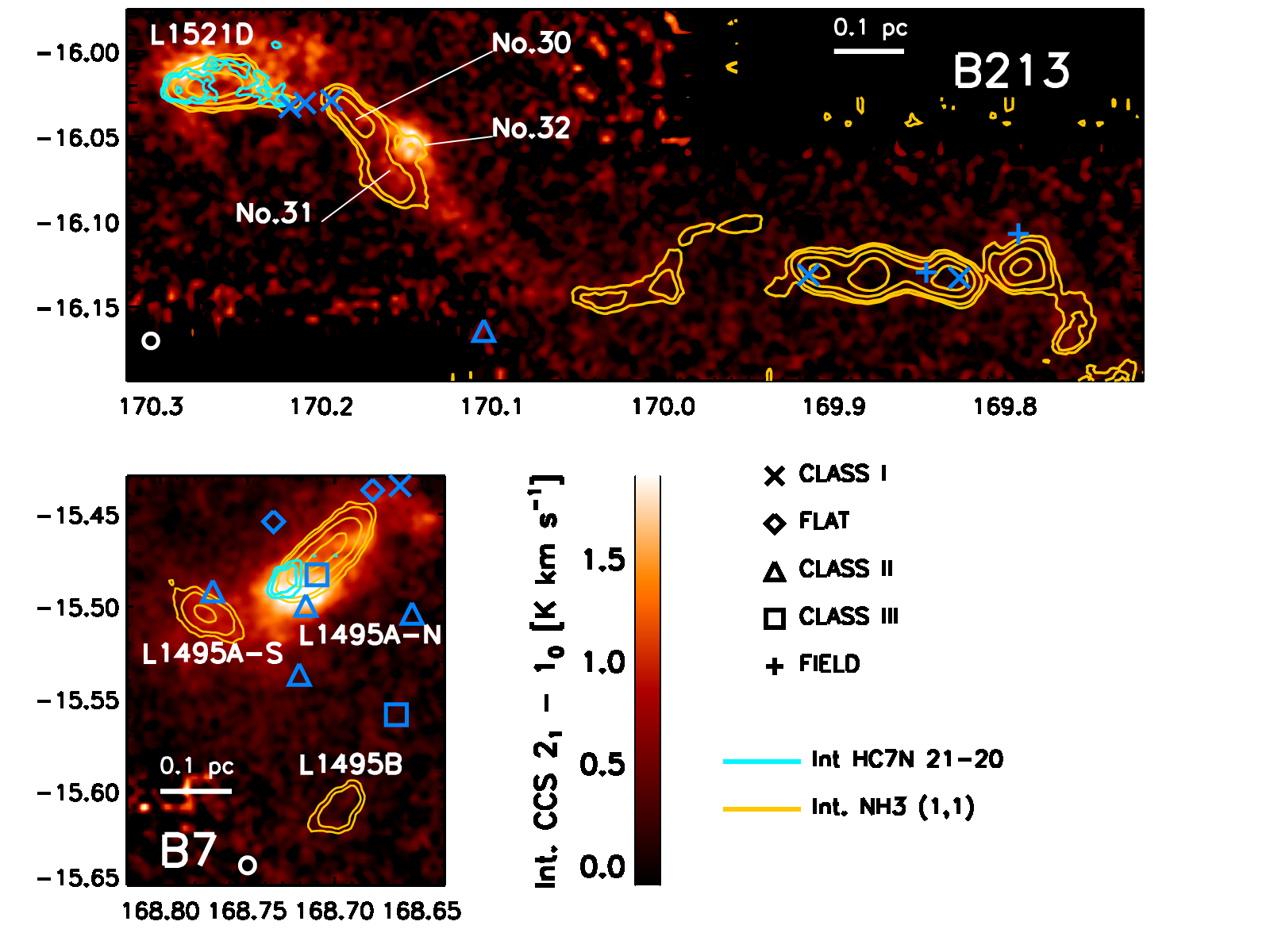}
\caption{{Images of integrated CCS 2$_1$ $-$ 1$_0$ (color), integrated NH$_3$ (1,1) (orange contours), HC$_7$N 21$-$20 (cyan contours), and positions of known protostars. The X-axis is Galactic longitude and the Y-axis is Galactic latitude. NH$_3$ contours are at integrated intensities of 1, 1.5, 3, 6, and 12 K km s$^{-1}$. HC$_7$N contours are at integrated intensities of 0.1, 0.15, 0.3, 0.6, 1.2 K km s$^{-1}$. The names of previously-studied cores discussed in this study are indicated in the images. For the dense cores discussed in this study but not having classical names, we used the IDs of NH$_3$ cores in \citet{seo15}}. The white circle in each image shows the beam size.}
\label{zoomin2}
\end{figure*}

We compare HC$_7$N $J$ = 21$-$20 (hereafter HC$_7$N 21$-$20) emission to CCS 2$_1$$-$1$_0$ and NH$_3$ (1,1) emission in Figure \ref{zoomin2}. HC$_7$N 21$-$20 emission is only detected in L1495A-N and L1521D at our $rms$ level of $\sim$85 mK. HC$_7$N 21$-$20 emission has a good correlation with strong CCS 2$_1$$-$1$_0$ emission of $>$1 K in both L1495A-N and L1521D. Similar to the CCS 2$_1$$-$1$_0$ emission, the HC$_7$N 21$-$20 intensity peaks do not agree with the NH$_3$ (1,1) intensity peaks or the dust continuum intensity peaks. Also, the HC$_7$N 21$-$20 emission in B213E has an arc-like shape that surrounds an NH$_3$ dense core. On the other hand, HC$_7$N 21$-$20 emission is not as spatially extended as CCS 2$_1$$-$1$_0$ emission, and the peak intensity of HC$_7$N 21$-$20 emission is less than 0.7 K in all regions, which is much weaker than CCS 2$_1$$-$1$_0$ intensity peaks ($>$2 K). The intensity peaks of HC$_7$N 21$-$20 emission are also slightly displaced from the CCS 2$_1$$-$1$_0$ intensity peaks. This may be due to different excitation conditions (E$_u^{\rm CCS}$/$k$ = 1.61 K while E$_u^{\rm HC_7N}$/$k$ = 12.5 K) and different chemical processes of CCS 2$_1$$-$1$_0$ and HC$_7$N 21$-$20, even though they are both carbon-chain molecules (see further discussion in \S4).


\subsection{Spectral Line Fitting}

\begin{figure*}[tb]
\centering
\includegraphics[angle=0,scale=1.]{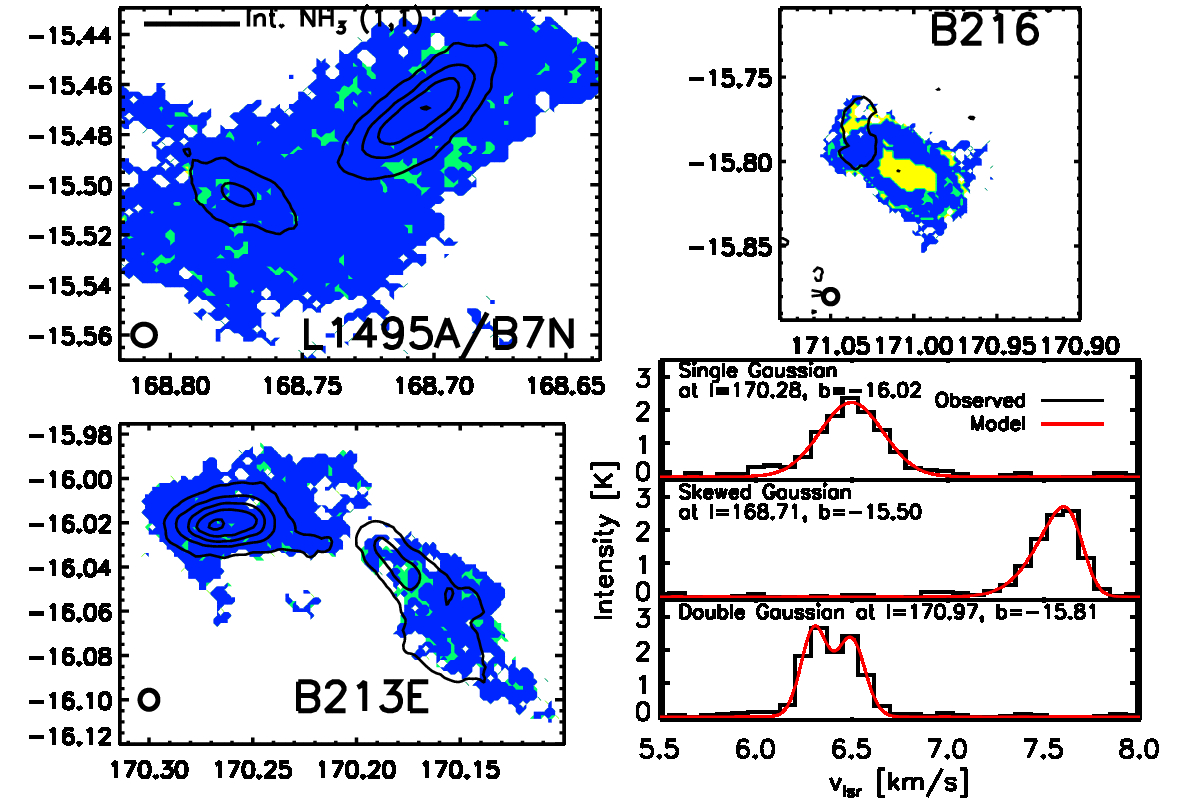}
\caption{Left and upper right panels: images of spectral fitting models (color) for CCS 2$_1$$-$1$_0$ lines with the integrated intensity of NH$_3$ (1,1) (black contours). Blue, cyan, and yellow denote areas in which single, skewed, and double Gaussian models, respectively, are used to fit spectral lines. NH$_3$ contours are at integrated intensities of 1, 1.5, 3, 6, and 12 K km s$^{-1}$. The black circle in each panel shows beam size. Single Gaussian fitting is forced unless $\chi^2$ is greater than 9 in the single Gaussian fitting. Left right panel: examples of spectra model fitting (red lines) to the observed lines (black lines). The top panel shows an example of single Gaussian fitting, the middle panel shows an example of skewed Gaussian fitting, and the bottom panel shows an example of double Gaussian fitting. The Galactic coordinates of the example spectra are written in each panel.  }
\label{fitting}
\end{figure*}

\begin{figure}[tb]
\centering
\includegraphics[angle=0,scale=0.9]{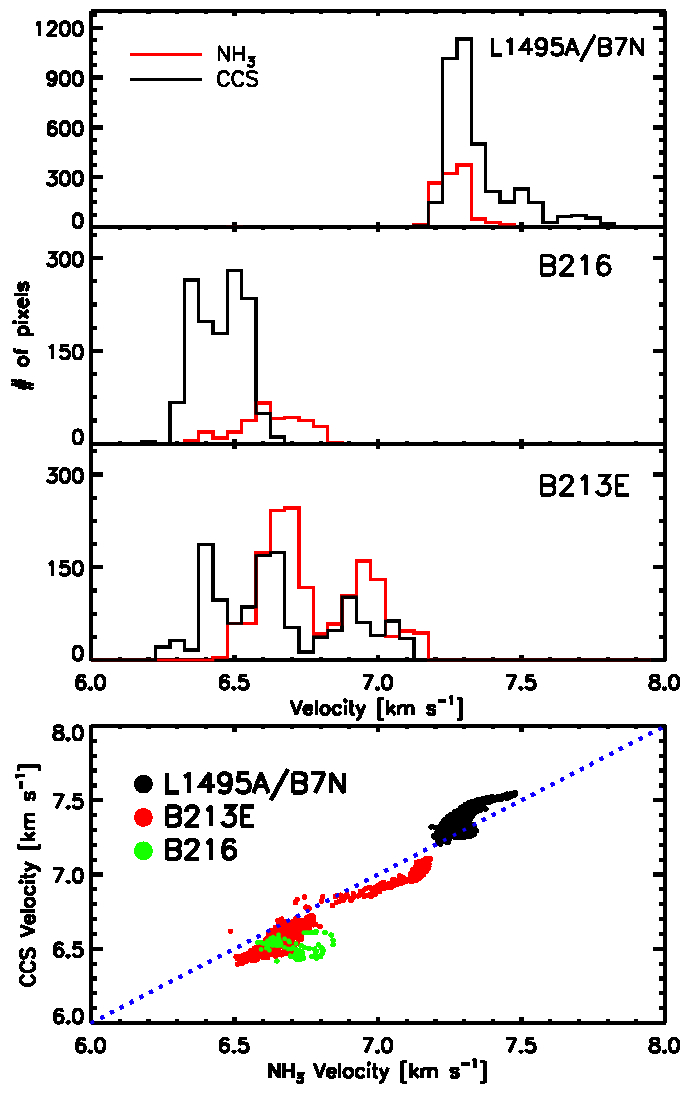}
\caption{{Top: Histograms of the LSR velocities in L1495A/B7N, B213E, and B216. Bottom: CCS 2$_1$$-$1$_0$ velocity vs. NH$_3$ velocity. Red histograms denote the LSR velocities of NH$_3$ emission, and black histograms denote the LSR velocities of CCS 2$_1$$-$1$_0$ emission in the top panel. In the bottom panel, each point corresponds to a pixel that has both CCS and NH$_3$ emission.}}
\label{hist_vlsr}
\end{figure}

Most of CCS 2$_1$$-$1$_0$ spectra in our survey are well described by a single Gaussian profile but some of CCS 2$_1$$-$1$_0$ spectra show skewed and double-peak profiles. Therefore, to understand the physical properties of observed CCS 2$_1$$-$1$_0$ structures in the L1495-B218 filaments, we fit CCS 2$_1$$-$1$_0$ lines with three different spectra models: single Gaussian, skewed Gaussian, and double Gaussian profiles. We use a least $\chi^2$ fit to the observed profile assuming the following model profiles:
\begin{eqnarray}
  \phi_{single}(v) &=& A \exp\left({(v-v_{lsr})^2\over 2\sigma^2} \right) \\
  \phi_{skewed}(v) &=& A \exp\left({(v-v_{lsr})^2\over2\sigma^2} \right)\left[1+{\rm erf}\left({\alpha (v-v_{lsr}) \over \sqrt{2}\sigma}\right) \right] \\
  \phi_{double}(v) &=& A_1 \exp\left({(v-v_{lsr,1})^2\over2\sigma^2_1} \right)+A_2 \exp\left({(v-v_{lsr,2})^2\over2\sigma^2_2} \right)
\end{eqnarray}
where $v$ is the velocity, $v_{lsr}$ is the LSR velocity of the line, $\sigma$ is the velocity dispersion, $A$ is the peak line intensity, $\alpha$ is the skewness, and the subscripts 1 \& 2 denote velocity components in the double Gaussian model. We found the $\chi^2$ values are typically lower with fits using the skewed Gaussian model relative to the single Gaussian since the skewed Gaussian model has one more degree of freedom in the line fitting. We used the single Gaussian model rather than the skewed Gaussian model if $\chi^2$ values from both models are less than 3$\sigma$.

Figure \ref{fitting} shows examples of observed CCS 2$_1$$-$1$_0$ lines and best-fit models. CCS 2$_1$$-$1$_0$ lines in all regions are fitted with single Gaussian profiles except in small portions of L1495A and B213E and the central part of B216. The number of spectra fitted with skewed Gaussian profiles is less than 5\% of the total number of spectra in every region. In B216, we found that there are two intensity peaks at 6.25 km s$^{-1}$, and 6.5 km s$^{-1}$ and the CCS 2$_1$$-$1$_0$ lines in the region are the better fit with double Gaussian profiles as indicated by reduced values of $\chi^2$.

Figure \ref{hist_vlsr} shows distributions of the LSR velocity of CCS 2$_1$$-$1$_0$ and NH$_3$ (1,1) emission in L1495A/B7N, B213E, and B216. In L1495A/B7N, the peaks of the LSR velocity distribution of CCS 2$_1$$-$1$_0$ emission and NH$_3$ emission show a good agreement with each other, which indicates that NH$_3$ (1,1) emission and CCS emission trace the same dense structures in this region. On the other hand, the LSR velocity range of CCS emission is considerably wider than that of NH$_3$ (1,1) emission with a long tail toward red-shifted velocity, which suggests CCS 2$_1$$-$1$_0$ emission traces extended structures in this area. In B216, the CCS 2$_1$$-$1$_0$ and NH$_3$ (1,1) emission overlap each other in the LSR velocity range from 6.2 km s$^{-1}$ to 6.6 km s$^{-1}$ but NH$_3$ (1,1) is more extended to 6.8 km s$^{-1}$. The histogram of the LSR velocity of CCS 2$_1$$-$1$_0$ emission shows two peaks at 6.25 km s$^{-1}$ and 6.5 km s$^{-1}$, while that of NH$_3$ emission shows a single peak at 6.6 km s$^{-1}$. Having CCS 2$_1$$-$1$_0$ emission at two velocities and a displacement between CCS 2$_1$$-$1$_0$ and NH$_3$ (1,1) intensity peaks suggest that B216 may be either a complex region with multiple dense structures at different LSR velocities or there may be a converging flow. However, we did not see any blue asymmetry in the HCN or HCO$^+$ 1$-$0 line profile indicating there is no significant converging motion. B213E includes four dense cores, and their LSR velocity ranges from 6.3 km s$^{-1}$ to 7.2 km s$^{-1}$. There are three peaks in the distribution of LSR velocity, and two of them coincide with NH$_3$ LSR velocity peaks. The peak at 6.9 km s$^{-1}$ is related of NH$_3$ core No.32 in the west of B213E. The other two peaks at 6.4 km s$^{-1}$ and 6.6 km s$^{-1}$ are related to L1521D. The LSR velocity ranges observed in CCS 2$_1$$-$1$_0$ and NH$_3$ (1,1) are quite similar.

Using the fitted profiles of CCS 2$_1$$-$1$_0$ lines, we estimated the column density of CCS under assumption of the optically thin emission. The CCS 2$_1$$-$1$_0$ emission in dense cores is generally optically thin \citep{wolkovitch97,rosolowsky08}. The estimated column density of CCS ranges from 1 $\times$ 10$^{12}$ cm$^{-2}$ to 3 $\times$ 10$^{13}$ cm$^{-2}$ when we assume the excitation temperature is same at the kinetic temperature. The median column density is 7 $\times$ 10$^{12}$ cm$^{-2}$. If we assume that the excitation temperature is 6 K, which is the excitation temperature found in TMC1, the column density is in the range of 1 $\times$ 10$^{12}$ $-$ 4 $\times$ 10$^{13}$ cm$^{-2}$. This column density range is similar to the CCS column densities found in other dense cores and Bok globules \citep[e.g.,][]{suzuki92, wolkovitch97, marka12}. Using a Monte-Carlo radiative transfer calculator, we found that the optical depth of CCS 2$_1$$-$1$_0$ for this column-density range is $<$0.5 with the median optical depth of 0.1, which is consistent with the optically-thin assumption.

We also fitted HC$_7$N 21$-$20 and evaluated the column density of HC$_7$N under the optically-thin limit. The profiles are well fitted with single Gaussian profiles and the column density ranges 1 $-$ 5 $\times$ 10$^{11}$ cm$^{-2}$ when we assume that the excitation temperature is same as the ammonia kinetic temperature $\sim$10 K. The HC$_7$N column density is roughly a factor of two larger if the excitation temperature is 4 K.

\subsection{Kinematics}

\subsubsection{L1495A-N}

\begin{figure*}[tb]
\includegraphics[angle=0,scale=0.85]{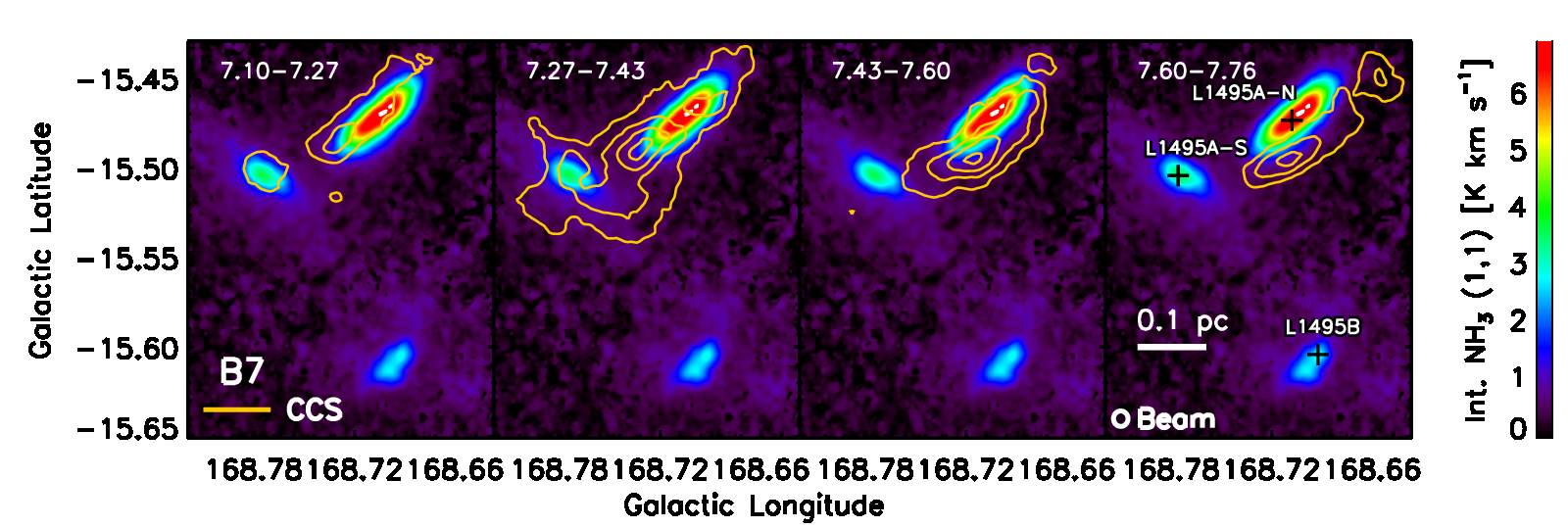}
\caption{{Channel map of CCS 2$_1$$-$1$_0$ (contours) and integrated intensity of NH$_3$ (1,1) (color) in L1495/B7. CCS 2$_1$$-$1$_0$ contours start at 0.1 K km s$^{-1}$ and increase in steps of 0.1 K km s$^{-1}$. The velocity range for each channel map is given at the top of each panel in units of km s$^{-1}$. Names of the dense cores are written in the last panel.}}
\label{l1495a_channel}
\end{figure*}

We probe the kinematics of dense core L1495A-N using spectral maps of CCS 2$_1$$-$1$_0$ and NH$_3$ (1,1) and investigate the motions of filaments and dense cores traced by the CCS 2$_1$$-$1$_0$ emission. Figure \ref{l1495a_channel} presents channel maps of the CCS 2$_1$$-$1$_0$ emission along with the integrated intensity of NH$_3$ (1,1) in the B7 region. The LSR velocities of NH$_3$ dense cores are 7.2 km s$^{-1}$ and 7.25 km s$^{-1}$ at their centers. The first panel in Figure \ref{l1495a_channel} shows the blue-shifted CCS emission relative to the NH$_3$ (1,1) emission in L1495A-N, whereas the two right panels present the red-shifted CCS 2$_1$$-$1$_0$ emission relative to the NH$_3$ (1,1) emission in L1495A-N. In the first panel, the blue-shifted CCS 2$_1$$-$1$_0$ emission is bright in the northeastern part of L1495A-N and elongated similar to the NH$_3$ core. With increasing LSR velocity, we find that CCS 2$_1$$-$1$_0$ emission moves toward the southeast direction, which indicates that L1495A-N has shear flows, or rotation, or converging flows.

\begin{figure}[tb]
\includegraphics[angle=0,scale=0.85]{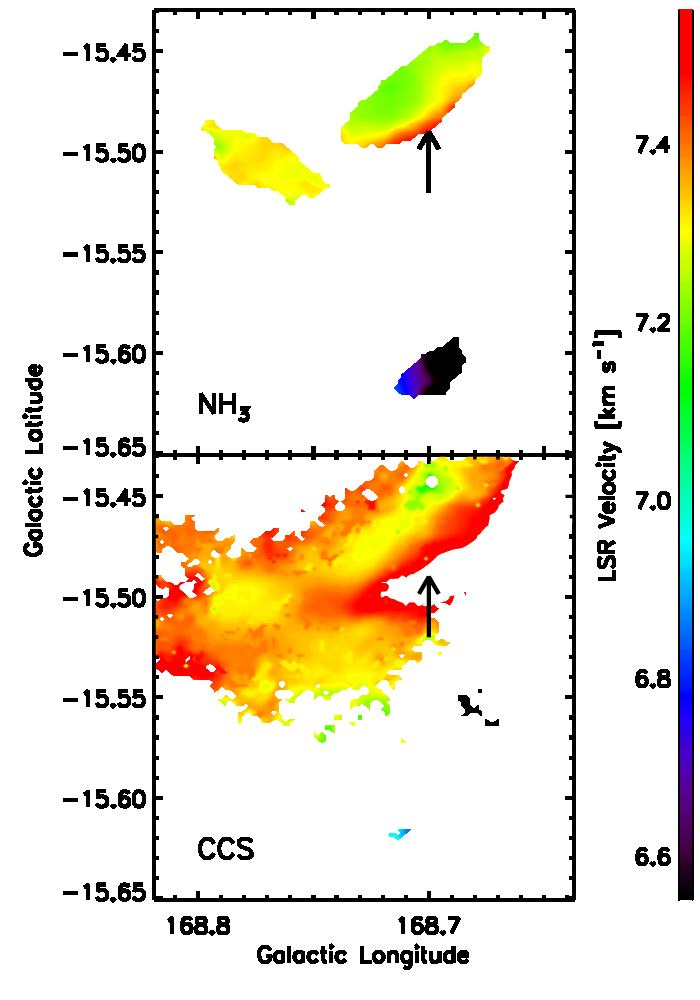}
\caption{{Maps of NH$_3$ (1,1) (top) and CCS 2$_1$$-$1$_0$ (bottom) LSR velocities in L1495. The black arrow points to the steep transition in the LSR velocity within L1495A-N.}}
\label{l1495a_vlsr}
\end{figure}

We investigate the flow of gas using the LSR velocity maps of CCS 2$_1$$-$1$_0$ and NH$_3$ (1,1) (Figure \ref{l1495a_vlsr}). We find a steep transition in the LSR velocities of both CCS 2$_1$$-$1$_0$ and NH$_3$ (1,1) emission within L1495A-N (pointed with black arrow). The transition shows a velocity shift of 0.3 km s$^{-1}$ within the size of beam (0.022 pc) and the steepest velocity gradient of $\sim$14 km s$^{-1}$ pc$^{-1}$ in both NH$_3$ (1,1) and CCS 2$_1$$-$1$_0$, which is at least seven times larger than the typical velocity gradient measured for rotation ($\sim$2 km s$^{-1}$ pc$^{-1}$, \citealt{goodman93}). Also, the transition is parallel to the long axis of L1495A-N rather than its short axis, which is a rare case. Only L183 is reported to have the rotation axis the rotation axis parallel to its long axis \citep{kirk09}. However, the velocity gradient of L1495A-N is considerably larger than that of L183 (8.9 km s$^{-1}$ pc$^{-1}$). Thus, an unusually large gradient of the LSR velocity in L1495A-N may be due to a converging flow or a shear flow but not all due to rotation.

\begin{figure*}[tb]
\centering
\includegraphics[angle=0,scale=0.85]{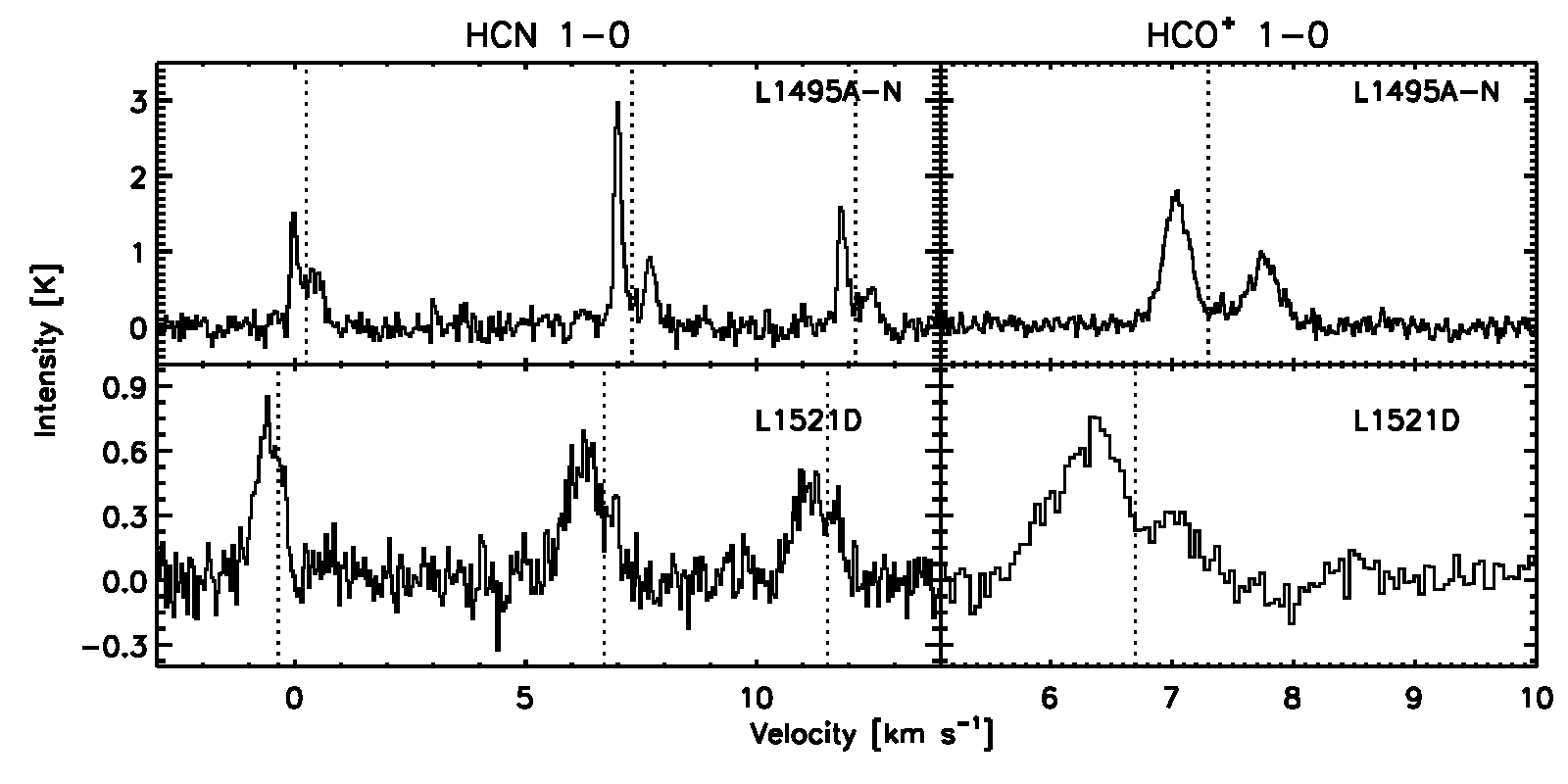}
\caption{{HCN 1$-$0 (left column) and HCO$^+$ 1$-$0 (right column) in L1495A-N (top) and L1521D (bottom). The HCN transition contains three hyperfine components $F$ = 1$-$0 (left), $F$ = 2$-$1 (center), and $F$ = 1$-$1 (right). The dashed vertical lines denote the LSR velocity of NH$_3$.}}
\label{HCNHCO}
\end{figure*}

\begin{figure*}[tb]
\centering
\includegraphics[angle=0,scale=0.6]{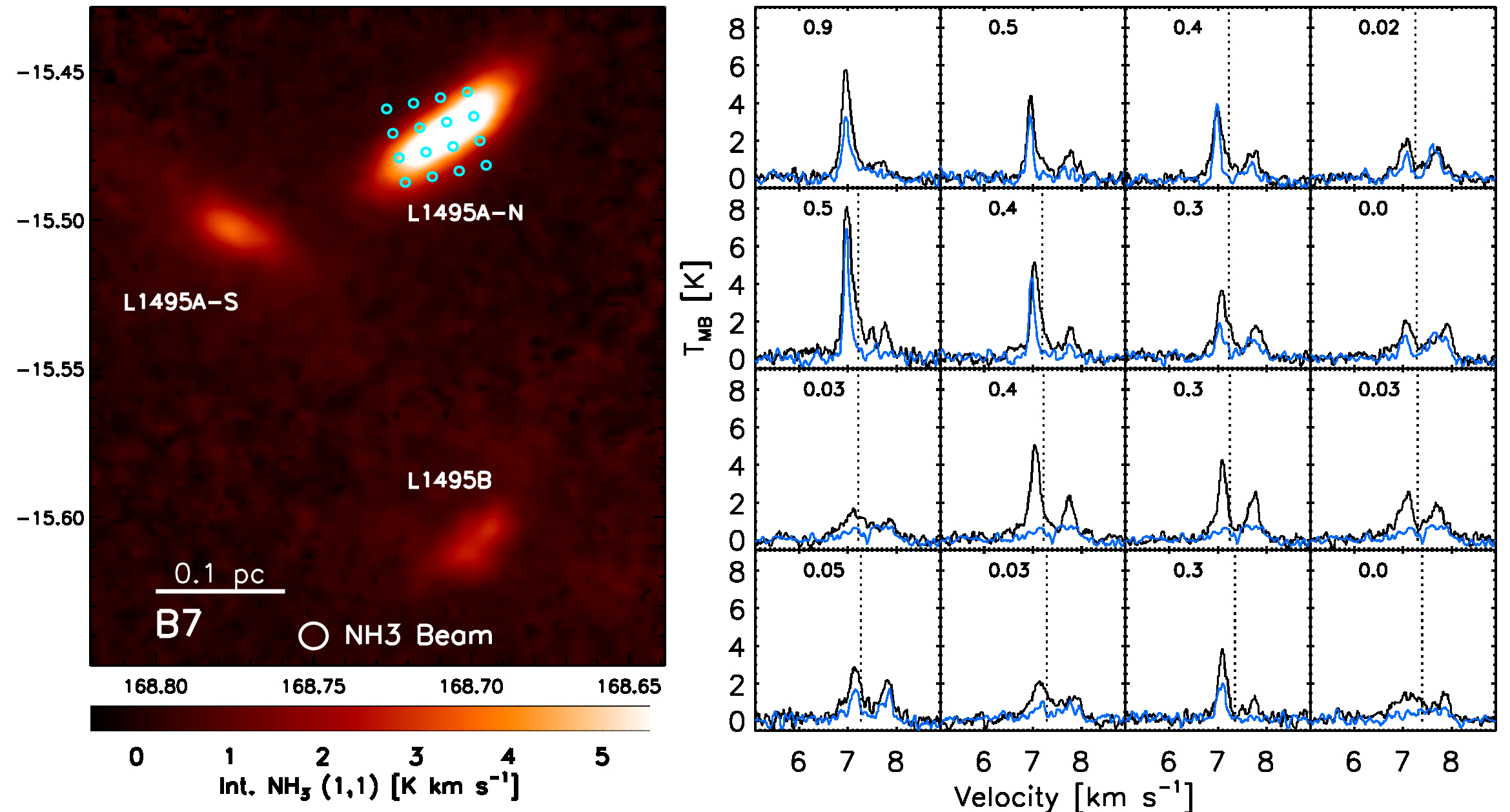}
\caption{{Integrated NH$_3$ (1,1) (left) and HCN \& HCO$^+$ 1$-$0 lines observed toward L1495A-N (right). Galactic coordinates are used in the left panel. The 16 small, cyan circles in the left panel denote the beams of Argus and the corresponding spectra are shown in the right panel with each box corresponding to an Argus beam. The position of the Argus observation is chosen for the 16 beams to cover the core center as well as its outskirts. HCO$^+$ 1$-$0 and HCN $J$ = 1$-$0, $F$ = 2$-$1 are shown as black and blues lines, respectively. The dotted vertical lines present the LSR velocities of the NH$_3$ emission. The speeds of converging motions evaluated using the two-layer model \citep{myers96} are indicated in each panel in units of km s$^{-1}$. Only the HCO$^+$ lines are used for evaluating the speed of converging-motions since the SNR of HCN is insufficient for accurate results.}}
\label{L1495AN_argus}
\end{figure*}

\begin{figure*}[tb]
\centering
\includegraphics[angle=0,scale=0.8]{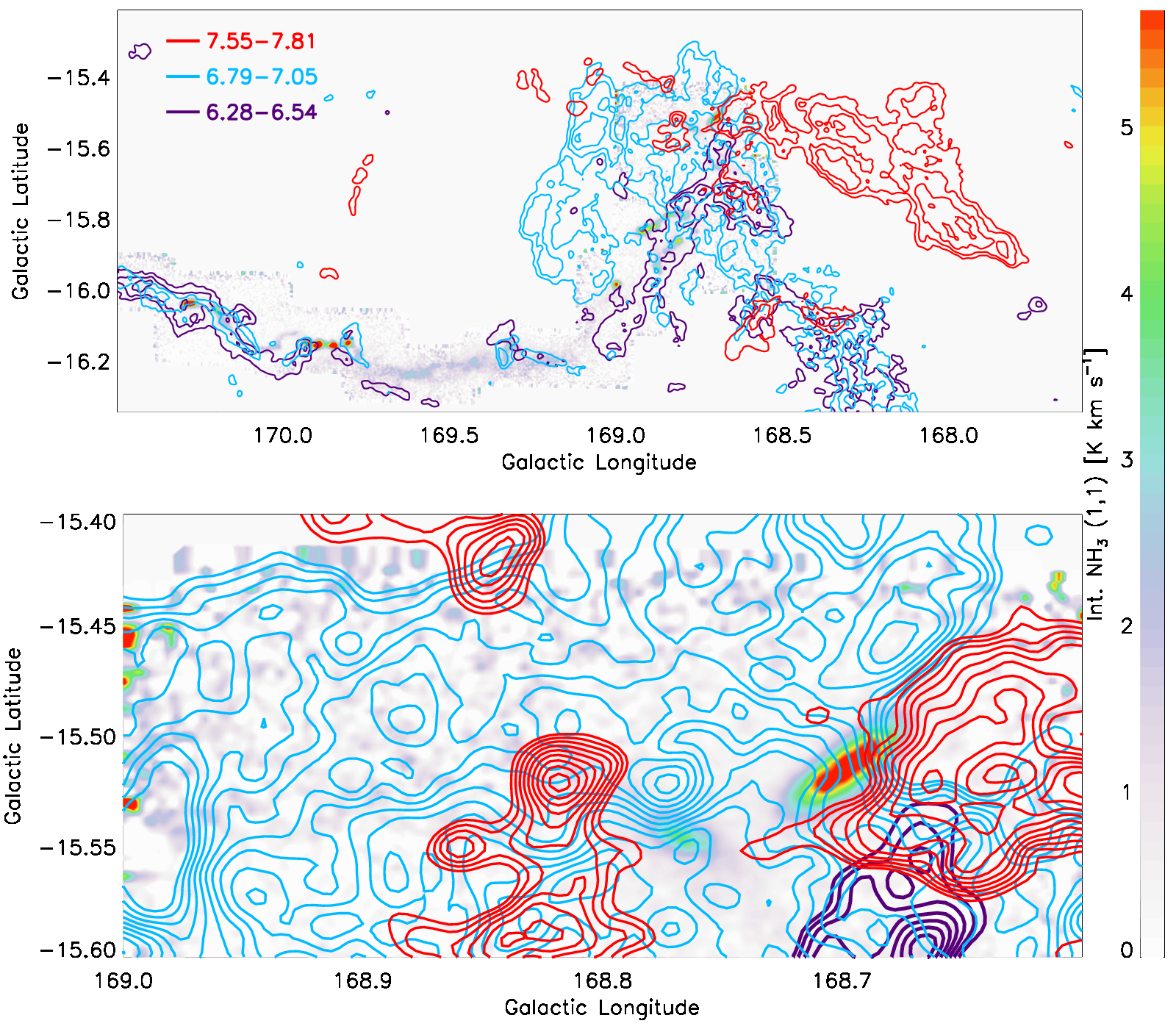}
\caption{{Channel maps of $^{13}$CO 1$-$0 (contours) and integrated intensity maps of NH$_3$ (1,1) (color) in the L1495-B218 filaments (top) and in L1495A/B7N (bottom). $^{13}$CO contours start at 0.44 K km s$^{-1}$ and increase in steps of 0.1 K km s$^{-1}$ for the top panel and in steps of 0.025 K km s$^{-1}$ for the bottom panel. The velocity range for each channel is written in different color in units of km s$^{-1}$.}}
\label{l1495a_channel_13co}
\end{figure*}
\begin{figure*}[tb]
\centering
\includegraphics[angle=0,scale=0.88]{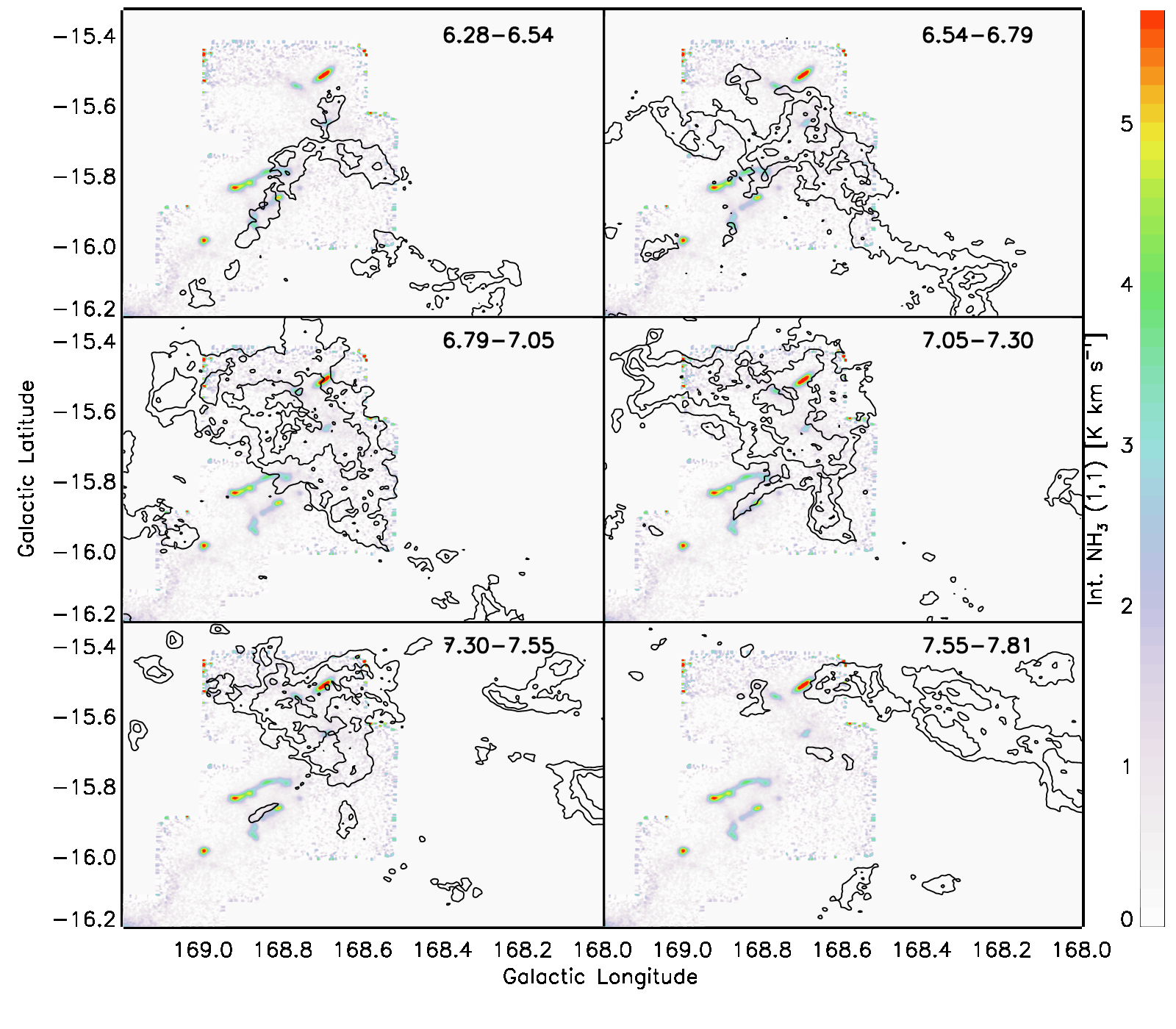}
\caption{{Channel maps of $^{13}$CO 1$-$0 (contours) and integrated intensity maps of NH$_3$ (1,1) (color) in L1495 and B10. $^{13}$CO contours start at 0.5 K km s$^{-1}$ and increase in steps of 0.13 K km s$^{-1}$. The velocity range for each channel is written in each panel in units of km s$^{-1}$.}}
\label{l1495a_channel_13co_new}
\end{figure*}
\begin{figure}[tb]
\centering
\includegraphics[angle=0,scale=0.5]{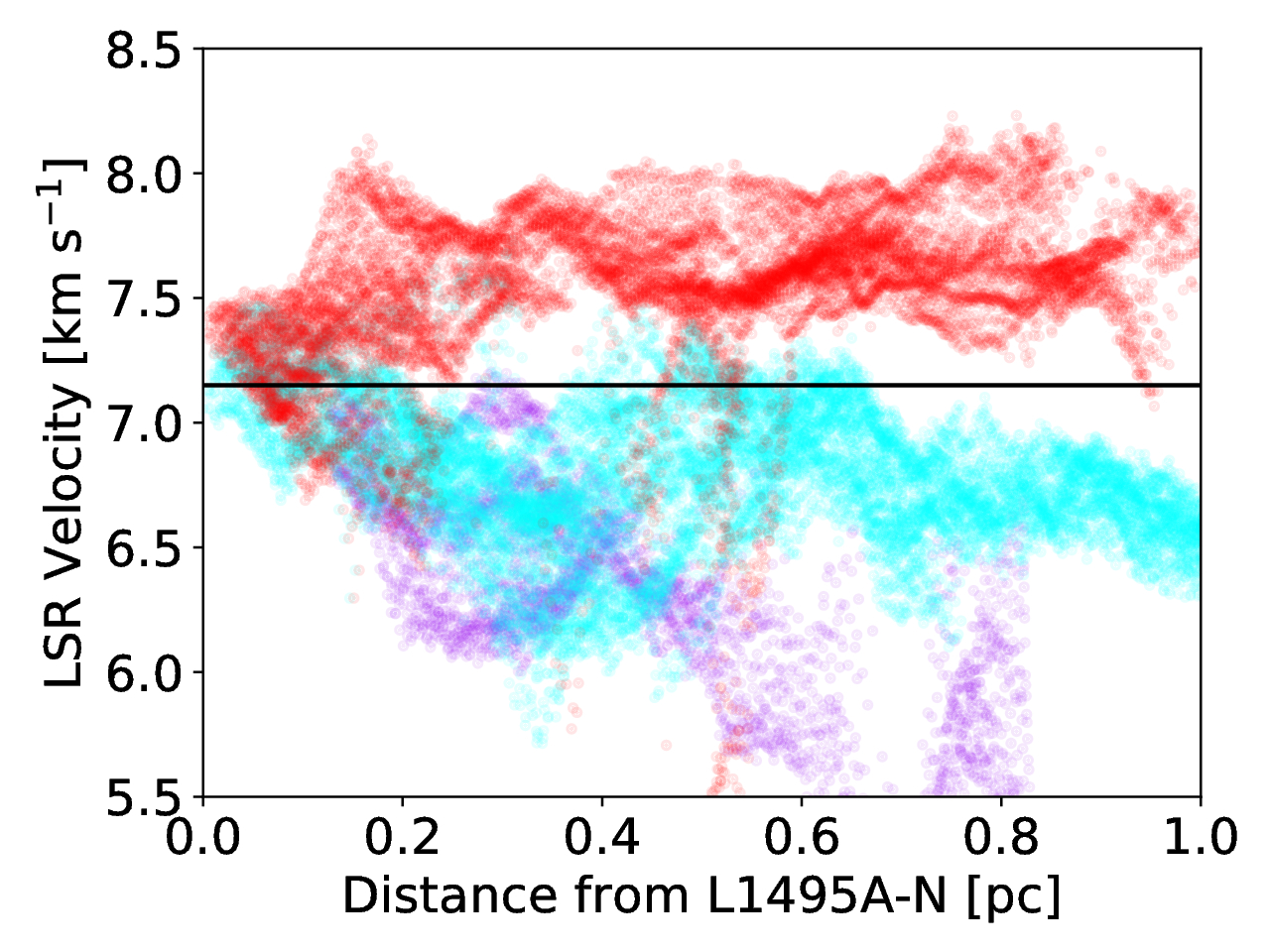}
\caption{{LSR velocities of the brightest $^{13}$CO lines along the three groups of CO structures from L1495A-N. The purple, cyan, and red dots represent spectra of Group I, II, and III, respectively, of the CO structures. The LSR velocity is evaluated by fitting a single Gaussian profile to the brightest component in a spectrum. The black horizontal line indicates the LSR velocity of L1495A-N at its center.}}
\label{velocity_gradient}
\end{figure}

To check whether the flow in L1495A-N is a converging flow, we observed L1495A-N in HCN $J$ = 1$-$0 and HCO$^+$ $J$ = 1$-$0 using the 12m ARO and 100m GBT. The beam of the 12m ARO covers the majority part of L1495A-N, thus the HCN 1$-$0 and HCO$^+$ 1$-$0 observations using the 12m ARO show the overall internal dynamics of L1495A-N. We found that L1495A-N has strong blue asymmetric profiles in both HCO$^+$ and HCN $J$ = 1$-$0 (Figures \ref{HCNHCO} and \ref{L1495AN_argus}), while optically thin CCS 2$_1$$-$1$_0$ and NH$_3$ (1,1) lines are well-fitted by a single Gaussian profile. To see whether the blue asymmetric profiles are due to self-absorption of a converging motion, we estimated the optical depths of HCN and HCO$^+$ 1$-$0 lines through our chemical model followed by LTE-radiative-transfer model \citep{majumdar17} and found their optical depths to be larger than 2 in L1495A-N. Also, $^{13}$CO 1$-$0 has two peaks at 6 km s$^{-1}$ and 7.2 km s$^{-1}$, which do not coincide the two intensity peaks of HCN and HCO$^+$ 1$-$0. These results indicate that the blue asymmetric profiles of HCN and HCO$^+$ 1$-$0 are not due to two different velocity components along the line of sight but are the results of self-absorption, which is often observed in collapsing dense cores \citep[e.g.,][]{lee99,sohn07}.

All hyperfine components of HCN $J$ = 1$-$0 observed using the ARO show blue asymmetric line profiles. The converging-flow velocities estimated using a simple two-layer model (Equation 9 in \citealt{myers96}) are 0.41 km s$^{-1}$ for $F$ = 2$-$1, 0.17 km s$^{-1}$ for $F$ = 1$-$1, and 0.10 km s$^{-1}$ for $F$ = 0$-$1. The converging-flow speed measured in HCO$^+$ 1$-$0 is 0.83 km s$^{-1}$, which is roughly four times larger than the sound speed. This converging-flow speed is higher than the maximum infall speed in a similarity solution of a gravitationally collapsing core \citep{larson69,penston69} and is similar to that formed in the collapse of dense core triggered by a supersonic converging flow \citep{gong09}.

Sparsely-mapped HCN and HCO$^+$ using Argus on the GBT shows the dynamics of L1495A-N at higher angular resolution (Figure \ref{L1495AN_argus}). While the 12m ARO data show overall dynamics of L1495A-N, the GBT data show spatially-resolved dynamics within L1495A-N and its surroundings since the beam size of GBT is much smaller than L1495A-N. Dense cores mapped previously in HCN 1$-$0 and HCO$^+$ 1$-$0, for example, L1544, L694-2, and L1197, all have strongest emission and largest infall speed (converging-flow speed due to gravitational collapse) at or near the core center and show a roughly azimuthally symmetric distribution of emission. On the contrary, L1495A-N has highly asymmetric distributions of HCN and HCO$^+$ intensities and converging-flow speeds. The brightest emission from HCN and HCO$^+$ is found on the northeast outskirts of the core rather than its center. Moreover, the converging-flow speeds are also the largest at northeast outskirts (0.9 km s$^{-1}$ or Mach 4.5), while the west side of L1495A-N shows almost no converging flow ($<$0.05 km s$^{-1}$). The quasi-static collapse models of a gas sphere and filaments typically have the maximum infall velocity less than Mach 4, and the fastest infall occurs very close to the core center or the axis of the filament ($<$0.01 pc) at the final stage of the collapse \citep{foster93,myers05,gong09,seo13,burge16}. The high converging-flow speed only at the outskirts and the large degree of asymmetry seen in L1495A-N do not agree with the quasi-static collapse models and suggest that large-scale gas flows are likely driving the dynamics of L1495A-N.

To see whether the converging flow observed in L1495A-N in HCN 1$-$0 and HCO$^+$ 1$-$0 is connected with a large-scale flow, we examine $^{13}$CO 1$-$0 integrated intensity and channel maps along with the integrated intensity map of NH$_3$ (1,1) in Figure \ref{l1495a_channel_13co}. The $^{13}$CO 1$-$0 is observed using the 14m FCRAO with a spatial resolution of 47$''$ and a velocity resolution of 0.256 km s$^{-1}$ \citep{goldsmith08}. We found that there are roughly three different groups of $^{13}$CO structures radiating from L1495 in the integrated intensity map and found in the channel maps that the hub region may be a place where the three groups of$^{13}$CO structuresare colliding or converging (upper panel of Figure \ref{l1495a_channel_13co}). In Figure \ref{l1495a_channel_13co}, we show the three groups of $^{13}$CO structures with the velocity ranges chosen to emphasize each group of $^{13}$CO structures. One group of $^{13}$CO structures (Group I at 6.28$-$6.54 km s$^{-1}$) stretches from the hub to B10, B213, \& B216 and forms the L1495-B218 filaments. Another group of $^{13}$CO structures (Group II at 6.79$-$7.05 km s$^{-1}$) is distributed from the hub and extends to a southwest direction. The last group of $^{13}$CO structures (Group III at 7.55$-$7.81 km s$^{-1}$) stretches from the hub to the west direction. We also present channel maps in a full velocity range from 6.28 km s $^{-1}$ to 7.81 km s$^{-1}$ in Figure \ref{l1495a_channel_13co_new} to probe the hub and the three groups in the Position-Position-Velocity (PPV) space. The $^{13}$CO 1$-$0 emission from 7.05 km s$^{-1}$ to 7.55 km s$^{-1}$ is the CO emission from the hub. Group III connects to the hub from red-shifted velocity and has the velocity gradient of roughly -0.8 km s$^{-1}$ pc$^{-1}$ from the Group I to the hub. On the other hand, Groups I and II are at blue-shifted velocities and bifurcate from the hub with the velocity gradients of 0.6 km s$^{-1}$ pc$^{-1}$ and 2 km s$^{-1}$ pc$^{-1}$, respectively. The velocity gradients are the median values of the velocity gradients measured following 1 pc distance along the three groups of $^{13}$CO structures from L1495A-N  (Figure \ref{velocity_gradient}). Groups II and III seem to be converging toward the same location in the hub, while Group I connects to a position slightly to the south. The fact that the velocity gradients of the three groups are toward the hub with signatures of converging motions in the HCN 1$-$0 and HCO$^+$ 1$-$0 observations and that the hub is the most massive and gravitational unstable part of the region based on a virial calculation suggest that the velocity gradients of the three groups are likely due to accretion of gas on the hub along the filaments \citep{seo15}.

We found that L1495A-N is embedded within the hub in PPV space and exactly located where Groups II and III interface with each other in a zoomed view of L1495A-N with $^{13}$CO channel maps (bottom panel of Figure \ref{l1495a_channel_13co}). The LSR velocity of $^{13}$CO 1$-$0 emission also connects seamlessly with the LSR velocity of NH$_3$ and CCS emission and the difference of the LSR velocities between the Groups II and III is consistent with the speed of converging gas flows measured in HCO$^+$, which suggests that L1495A-N and the large-scale $^{13}$CO gas are connected in the PPV space. Having a coherent gas flow from parsec to sub-parsec scale suggests that the formation and evolution of the L1495A-N core are not determined solely by local dynamics but may be driven by large-scale gas flows.

\subsubsection{L1521D \& B213E}

\begin{figure*}[tb]
\centering
\includegraphics[angle=0,scale=0.88]{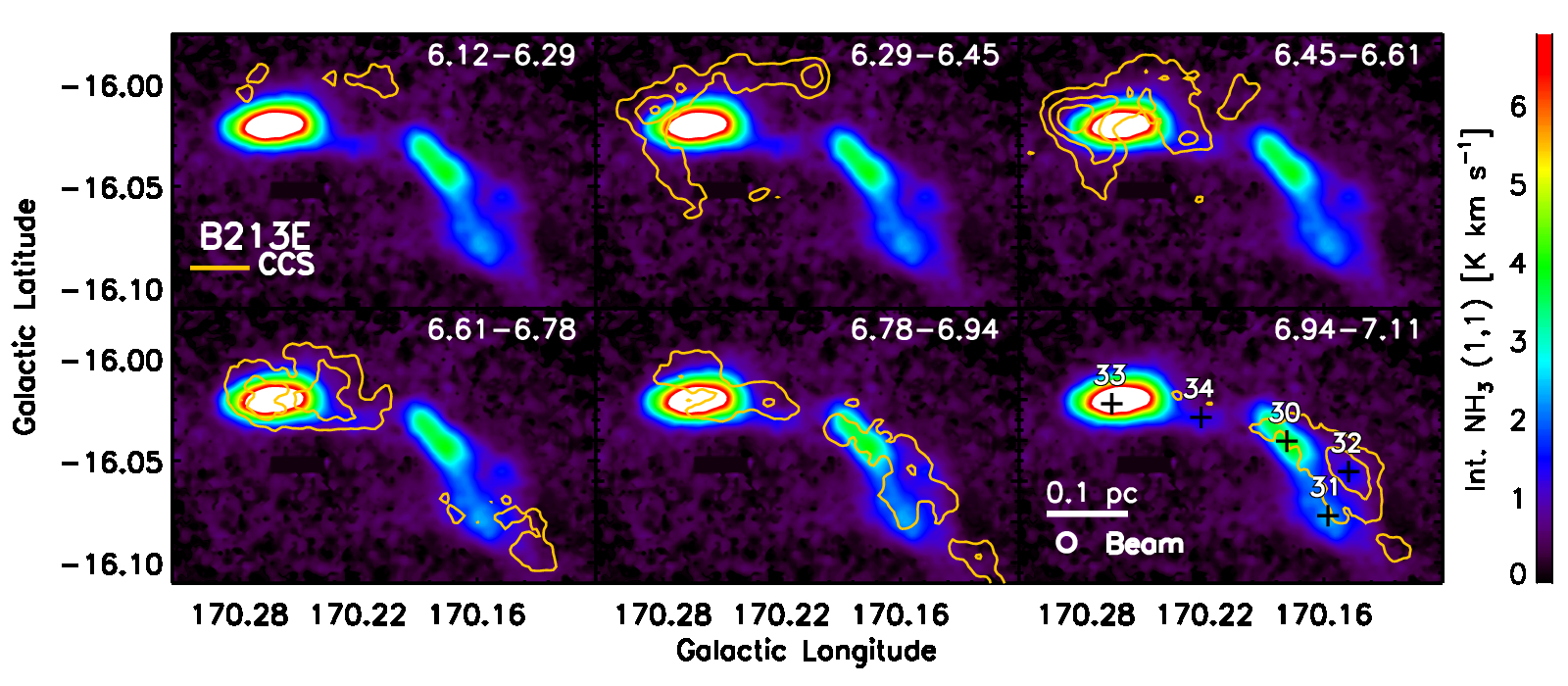}
\caption{{Channel map of CCS (contours) and integrated intensity of NH$_3$ (1,1) (color) in B213E. CCS contours start from 0.1 K km s$^{-1}$ and increase in steps of 0.1 K km s$^{-1}$. The velocity range for each channel map is written in each panel in km s$^{-1}$. The IDs of NH$_3$ dense cores are marked in the last panel.}}
\label{b213e_channel}
\end{figure*}

We probe the kinematics of the B213E region containing dense cores L1521D, No.30, No.31, and No.32 using channel maps of CCS and NH$_3$ (Figure \ref{b213e_channel}). The NH$_3$ LSR velocity of L1521D is 6.75 km s$^{-1}$ at its center. The blue-shifted CCS emission is bright and covers the outskirts of L1521D except for the south side of the core (upper three panels). The red-shifted CCS emission is bright at the center and the west side of L1521D (the bottom three panels). The last two panels in Figure \ref{b213e_channel} show CCS emission near the west side of B213E where there are three NH$_3$ dense cores \citep{seo15}. It shows that the CCS intensity peak coincides with the NH$_3$ intensity peak of the NH$_3$ core No.32 in the channel map.

\begin{figure}[tb]
\includegraphics[angle=0,scale=1.]{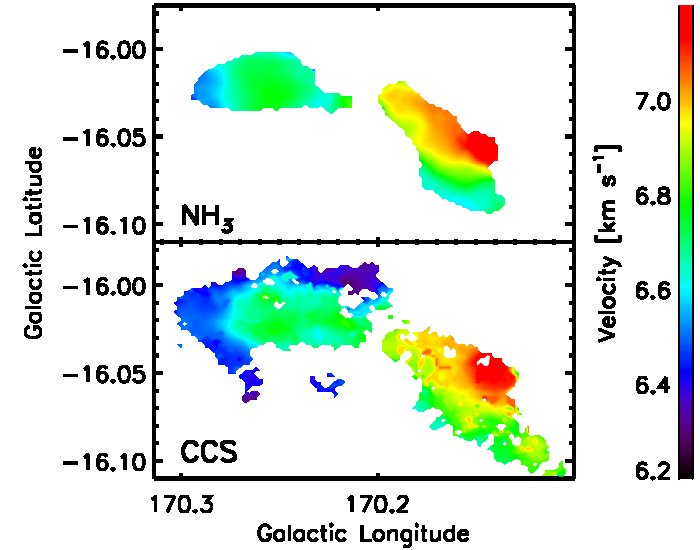}
\caption{{Maps of the NH$_3$ (top) and CCS (bottom) LSR velocities in B213E.}}
\label{b213e_vlsr}
\end{figure}

\begin{figure*}[tb]
\centering
\includegraphics[angle=0,scale=0.6]{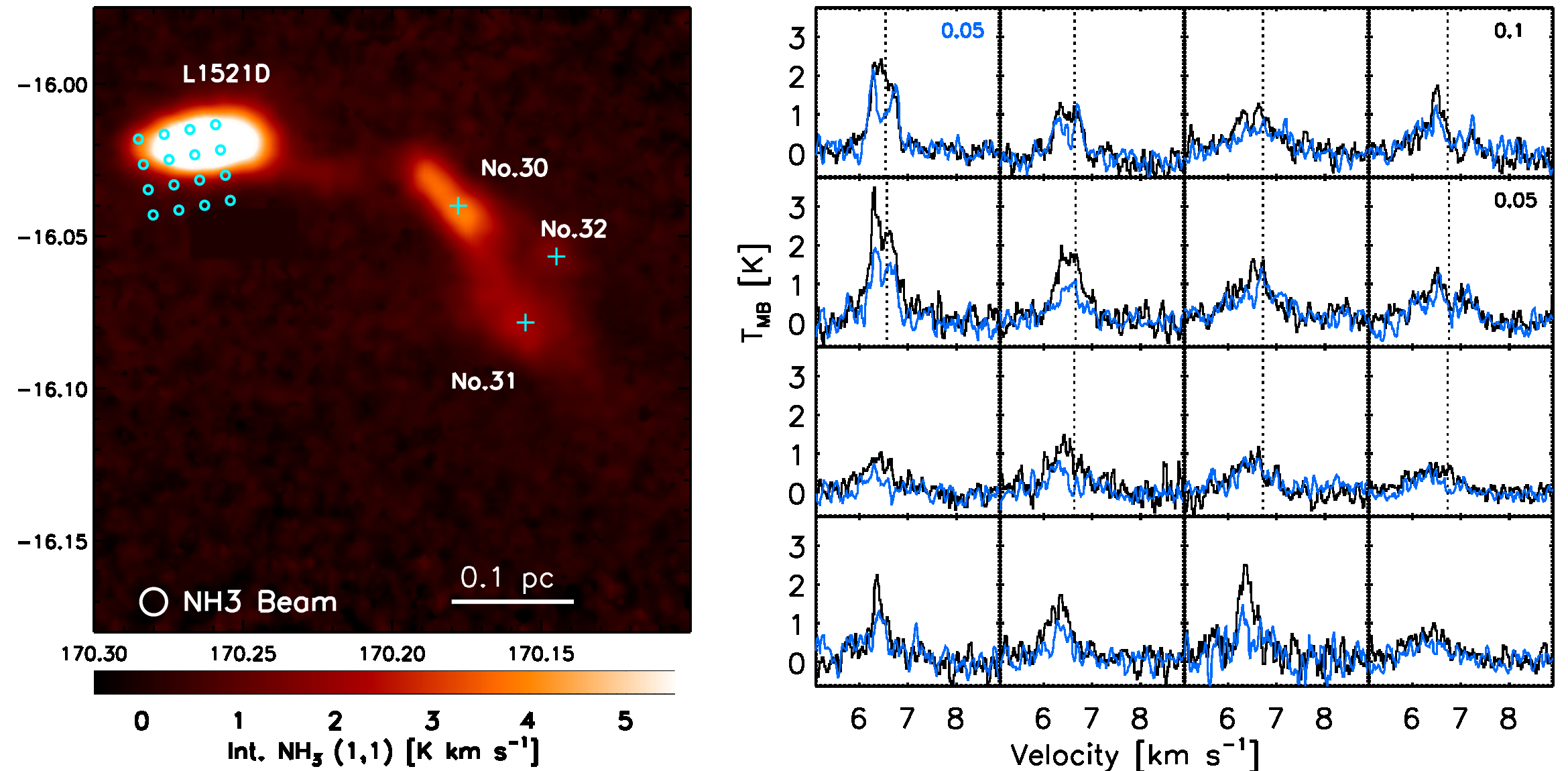}
\caption{{Integrated NH$_3$ (1,1) (left) and HCN \& HCO$^+$ 1$-$0 lines observed toward L1521D (right). The Galactic coordinates are used in the left panel. The 16 small, cyan circles in the left panel denote the beams of Argus and the spectra are shown in the right panel. The position of the Argus observation is chosen for the 16 beams to cover the core center as well as its outskirts. The boxes in the right panel correspond to the Argus beams in the left panel. HCO$^+$ 1$-$0 and HCN $J$ = 1$-$0, $F$ = 2$-$1 are shown as black and blues lines, respectively. The dotted vertical lines present the LSR velocities of the NH$_3$ emission. The speeds of converging motions evaluated using the two-layer model \citep{myers96} are written in the unit of km s$^{-1}$. The speeds evaluated using HCO$^+$  and HCN lines are written in black and blue, respectively.}}
\label{L1521D_argus}
\end{figure*}

\begin{figure*}[tb]
\centering
\includegraphics[angle=0,scale=0.88]{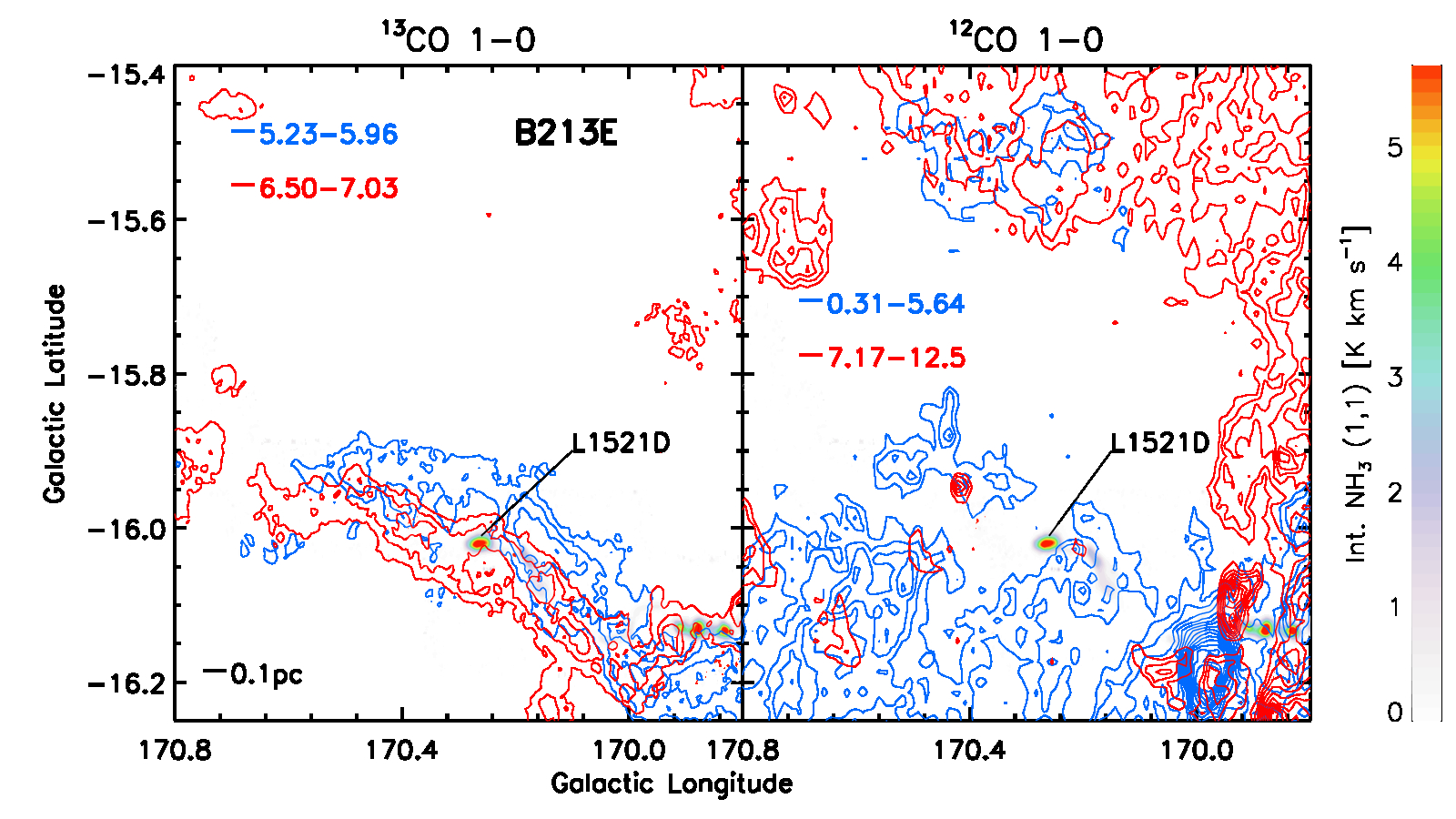}
\caption{{Channel maps of $^{13}$CO 1$-$0 (contours, left) and $^{12}$CO 1$-$0 (contours, right) overlaid on integrated intensity NH$_3$ (1,1) map in the B213E region. $^{13}$CO contours start at 0.75 K km s$^{-1}$ and increase in steps of 0.23 K km s$^{-1}$. $^{12}$CO contours start at 1.5 K km s$^{-1}$ and increase in steps of 0.46 K km s$^{-1}$. The velocity ranges (in km s$^{-1}$) are denoted by different colors and indicated in each panels. The velocity range for each channel is written in different color in the unit of km s$^{-1}$.}}
\label{b213e_channel_13co}
\end{figure*}

We further investigate the kinematics using the LSR velocity map of B213E in Figure \ref{b213e_vlsr}. In L1521D, the velocity gradient within L1521D is 3 km s$^{-1}$ pc$^{-1}$ in both NH$_3$ and CCS, which is similar to the average velocity gradient of fast rotating dense cores \citep{goodman93}. In B213E, the LSR velocity changes steeply from the dense core No.31 to the dense core No.32 with 13 km s$^{-1}$ pc$^{-1}$, which is a considerably larger gradient than a fast rotating core. The transition may be due to either the two cores are at different LSR velocities which overlap at their outskirts or due to a recent collision with each other.

In order to see whether L1521D and dense cores No.30, No.31, and No.32 have converging motionss, we observed them in HCN 1$-$0 and HCO$^+$ 1$-$0 using the 12m ARO and the 100m GBT with Argus. We found that only L1521D shows blue asymmetric profiles in HCO$^+$ 1$-$0 and HCN 1$-$0 (Figures \ref{HCNHCO}, \ref{L1521D_argus}). Using a radiative transfer calculation, we evaluated the optical depths of the observed HCN and HCO$^+$ 1$-$0 lines and found that they are $\geq$2, which is similar to the optical depths of other dense cores in HCN and HCO$^+$ 1$-$0 \citep{devries05,williams06,campbell16}. This confirms that the blue asymmetric profiles of HCN 1$-$0 and HCO$^+$ 1$-$0 in L1521D are the self-absorbed features due to a converging motion. The HCN $J$ = 1$-$0 emission of L1521D observed using the 12m ARO is relatively weak compared to HCO$^+$ $J$ = 1$-$0 emission (Figure \ref{HCNHCO}). On the other hand, HCO$^+$ $J$ = 1$-$0 from the 12m ARO has bright emission ($>$1 K) with a blue asymmetric, self-absorbed profile. The estimated converging-motion speed using HCO$^+$ 1$-$0 is 0.08 km s$^{-1}$ using the two-layer model \citep{myers96}, which is a bit less than half of the sound speed and substantially slower than the speeds measured in L1495A-N.

Observations toward L1521D using the GBT show similar internal dynamics with those found using the 12m ARO. The position of the observation is chosen to cover the outskirts and the inner region of L1521D. L1521D shows converging-motion signatures on both east and west sides with converging-motion speeds in the range 0.05 $-$ 0.1 km s$^{-1}$. HCN and HCO$^+$ 1$-$0 intensities are stronger on the east side of the core, similar to CCS emission, which may be due to an asymmetry in the chemistry or the excitation. The converging-motion speeds measured in HCN \& HCO$^+$ 1$-$0 are similar on both sides of the core, and no significant asymmetry in the internal dynamics is found. As we shall discuss in \S4, in terms of the kinematics, L1521D is similar to a slowly contracting core due to self-gravity. As for dense core No.32, mapping using the GBT shows complex line profiles and that there are at least three velocity components, but no suggestion of a significant converging motion.

To see whether the gas motions in L1521D are connected with large-scale flows, we investigate $^{13}$CO and $^{12}$CO channel maps together with NH$_3$ and dust structures in B213E in Figure \ref{b213e_channel_13co}. The velocity ranges of $^{13}$CO and $^{12}$CO channel maps are chosen to highlight the CO flows, so the red-shifted and blue-shifted components are not isolated structures but continuous flows in the PPV space. In the region close to L1521D, the $^{13}$CO 1$-$0 emission is red-shifted in the south of L1521D and is blue-shifted in the north of L1521D. On the other hand, on a large scale, the $^{12}$CO 1$-$0 emission is red-shifted in the northwest of L1521D and is blue-shifted in the south of L1521D. The CCS and NH$_3$ flows in L1521D do not have the same direction as the $^{12}$CO flows but are seamlessly connected with $^{13}$CO in the PPV space. Hence, the gas flow in L1521D is possibly a local flow restricted by the filaments but not connected to the large-scale flows that follow the striations seen in dust continuum emission \citep{palmeirim13}. To the west side of B213E where three NH$_3$ cores (No.30, No.31, No.32) are located, the LSR velocity of CCS changes from the southeast to the northwest direction, a shift which is aligned with the LSR velocities of both $^{12}$CO and $^{13}$CO.

\subsubsection{B216}

\begin{figure*}[tb]
\centering
\includegraphics[angle=0,scale=0.88]{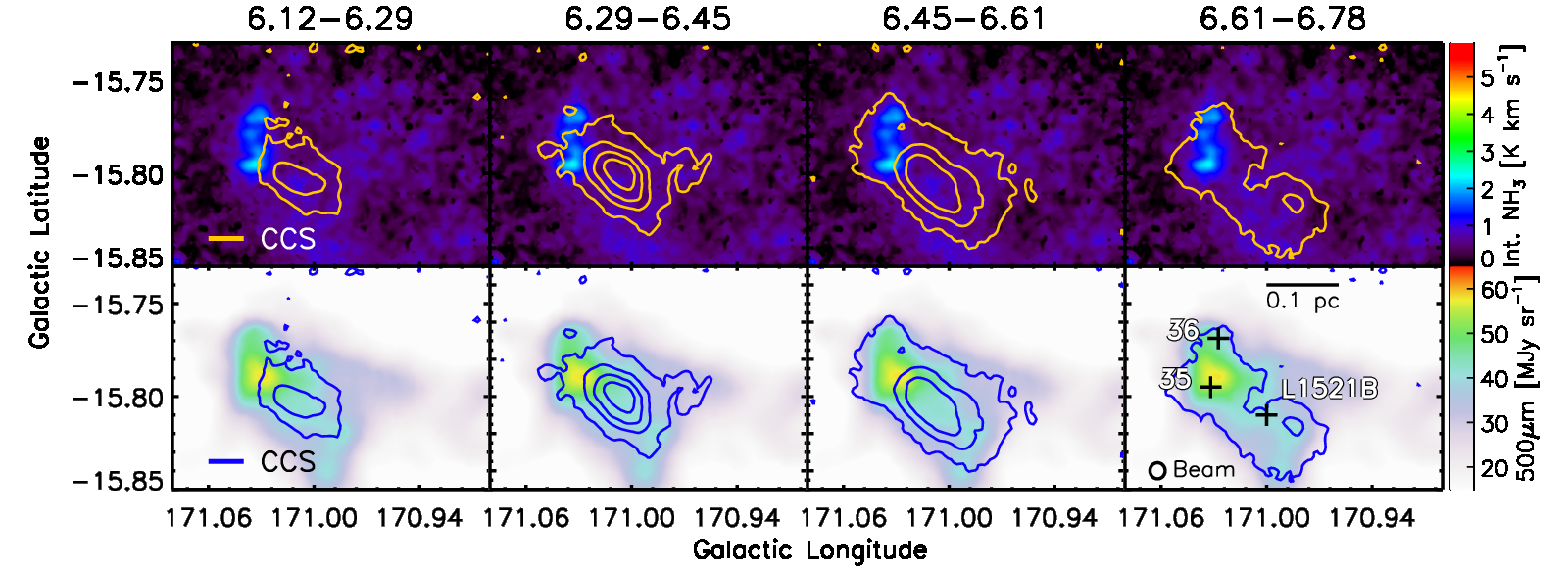}
\caption{Channel map of CCS (contours) and integrated intensity of NH$_3$ (1,1) (color) in B216. CCS contours start at 0.03 K km s$^{-1}$ and increase in steps of 0.11 K km s$^{-1}$. The velocity range in km s$^{-1}$ for each channel map is written above each column.}
\label{b216_channel}
\end{figure*}

We investigate the kinematics in the B216 region using the channel maps of CCS compared to the integrated NH$_3$ (1,1) and dust continuum at 500$\mu$m (Figure \ref{b216_channel}). The CCS emission in B216 spans 6.10 km s$^{-1}$ to 6.87 km s$^{-1}$. The CCS intensity peak does not coincide with NH$_3$ or dust continuum intensity peaks but is centered 0.08 pc to the southwest from the NH$_3$ peaks and dust continuum peaks. The two NH$_3$ dense cores, No.35 and No.36, have LSR velocities of 6.66 km s$^{-1}$ and 6.76 km s$^{-1}$. CCS emission only at 6.6 $-$ 6.8 km s$^{-1}$ is spatially associated with NH$_3$ cores and dust continuum peak. The CCS peak (also known as L1521B) was previously observed by \citet{hirota04} as an example of a CCS-bright young core. Our observations indicate that CCS does not trace the dense core in this region.

\section{DISCUSSION}

\subsection{CCS and HC$_7$N Chemistry in L1495-B218 Filaments}

\subsubsection{Our observations and previous studies}

The CCS-bright ``core"  predicted by \cite{suzuki92} is rarely found. In surveys toward the Taurus molecular cloud \citep{suzuki92, hirota09}, CCS is detected in 18 out of 29 cores, but only dense core L1521E is confirmed to have a CCS emission peak coinciding with the dust continuum peak \citep{hirota02,tafalla04}, which indicates that the lifetime of the CCS peak is at most 0.04 Myr (lifetime of a dense core is roughly 1 Myrs at the core mean density of $\bar{n}_{\rm H_2}$ = 10$^4$ cm$^{-3}$, \citealt{andre14}). The frequency of finding CCS-bright cores would be higher if Suzuki's chemical model were true for all cores since CCS is abundant up to 1 Myr in Suzuki's model.

We find that CCS is rarely bright at the center of the cores traced by NH$_3$ or dust continuum, but rather peaks on the core outskirts or outside the cores. Only one core (No.32) out 39 NH$_3$ cores along the L1495-B218 filaments has a good agreement among the dust, NH$_3$, and CCS intensity peaks. Thus, there are only two dense cores (No.32 \& L1521E) in the Taurus molecular cloud known to have coincident CCS and dust continuum intensity peaks. In particular, CCS emission in B216, which contains only young dense cores with no star formation activity, clearly shows that CCS intensity peaks do not coincide with NH$_3$ or dust continuum peaks (L1521B), even at an early evolutionary stage. This suggests that CCS is not a good tracer with which to trace core density structure.

We find that CCS emission is detected in both less-evolved regions (B216 and B211) and the more-evolved regions with star formation activity (L1495A/B7N and B213E), while we did not detect any significant CCS emission in B213W and B10 which seem to be at evolutionary stages slightly younger or similar to L1495A/B7N and B213E \citep{seo15}. Comparing the column density among the starless cores, we find that two more-condensed starless cores (e.g., L1521D and L1495A-N) have the brightest CCS and HC$_7$N emission while some other starless cores at the similar column density (e.g., starless cores in B213W) have no significant CCS and HC$_7$N detection at our noise level. We also find that the less-condensed starless cores and filaments in our survey (e.g., starless cores in B216 and filaments in B211) have CCS emission but no HC$_7$N emission. In the moderately-condensed starless cores in B10, we could not detect significant CCS or HC$_7$N emission. In summary, our CCS survey toward the L1495-B218 filaments shows that there is significant CCS emission in the less-condensed starless cores and filaments but also occasionally in more-condensed starless cores. This is not consistent with the prediction of the Suzuki's model which predicts that most of the starless cores in Taurus should have significant CCS emission.

\citet{aikawa01} made a chemical model of a dense core considering grain-surface chemistry and collapse dynamics. They demonstrated that the CCS abundance is sensitive more to the absolute age of the cores rather than to the gas density, suggesting that a more-condensed core can have a high CCS abundance if it is young. For example, L1544 and L1689B have similar densities and show signatures of converging-motion but L1689B has considerably brighter CCS emission than L1544. This may be because L1689B may be younger or has accreted surrounding gas more recently than L1544 \citep{lee03, sohn07, seo13}. While Aikawa's model suggests a possibility of having high CCS abundance in more-condensed starless cores, Aikawa's model still predicts considerably more CCS-bright cores than our observations and is relatively similar to Suzuki's model. This may be because the physical and chemical conditions of L1495-B218 filaments are different from Aikawa's model and their chemical network including both gas and grain surface chemistry is relatively simple. In this study, we evaluate chemical evolution in the L1495-B218 filaments using a state-of-the-art astrochemistry code using the observed physical conditions in Taurus.

\subsubsection{Chemistry of typical starless cores in the L1495-B218 filaments}

We modeled the chemistry in dense cores in Taurus using the state-of-the-art chemical code {\tt NAUTILUS} \citep{ruaud16}. It simulates a three-phase chemistry including gas-phase, grain-surface, and grain-mantle chemistry where both surface and the mantle are chemically active with possible exchanges between the different phases. Adsorption of gas-phase species on to grain surfaces, the thermal and non-thermal desorption of species from the grain surface into the gas phase, and species from the surface to mantle and mantle to the surface are the primary exchange processes. This chemical code considers the latest updates on various physico-chemical processes from the KIDA database\footnote{http://kida.obs.u-bordeaux1.fr/} which is more advanced as compared to previous studies reported in \citet{suzuki92} and \citet{aikawa01}. {\tt NAUTILUS} has been tested, benchmarked and applied to multiple cases such as dense cores \citep{vidal17,majumdar17b}; low-mass protostellar envelopes \citep{majumdar17,majumdar18b}, hot cores \citep{vidal18} and protoplanetary disks \citep{wakelam16}. The gas-phase chemistry of {\tt NAUTILUS} is based on the public chemical network kida.uva.2014 \citep{wakelam15} with updates for sulfur chemistry \citep{vidal17} and the chemistry of complex organics \citep{majumdar18a} from the KIDA database, whereas surface reactions are based on \citet{garrod08} with additional updates from \citet{ruaud16}, \citet{vidal17}, and \citet{majumdar18a}.  Using {\tt NAUTILUS}, we explored a wide range of parameters including density, temperature, visual extinction, and initial elemental abundances to model the chemistry in different environments. We limited excessive complexity of three-phase physio-chemical modeling by assuming that there is no inflow to the cores.

\begin{figure*}
\includegraphics[scale=1.15]{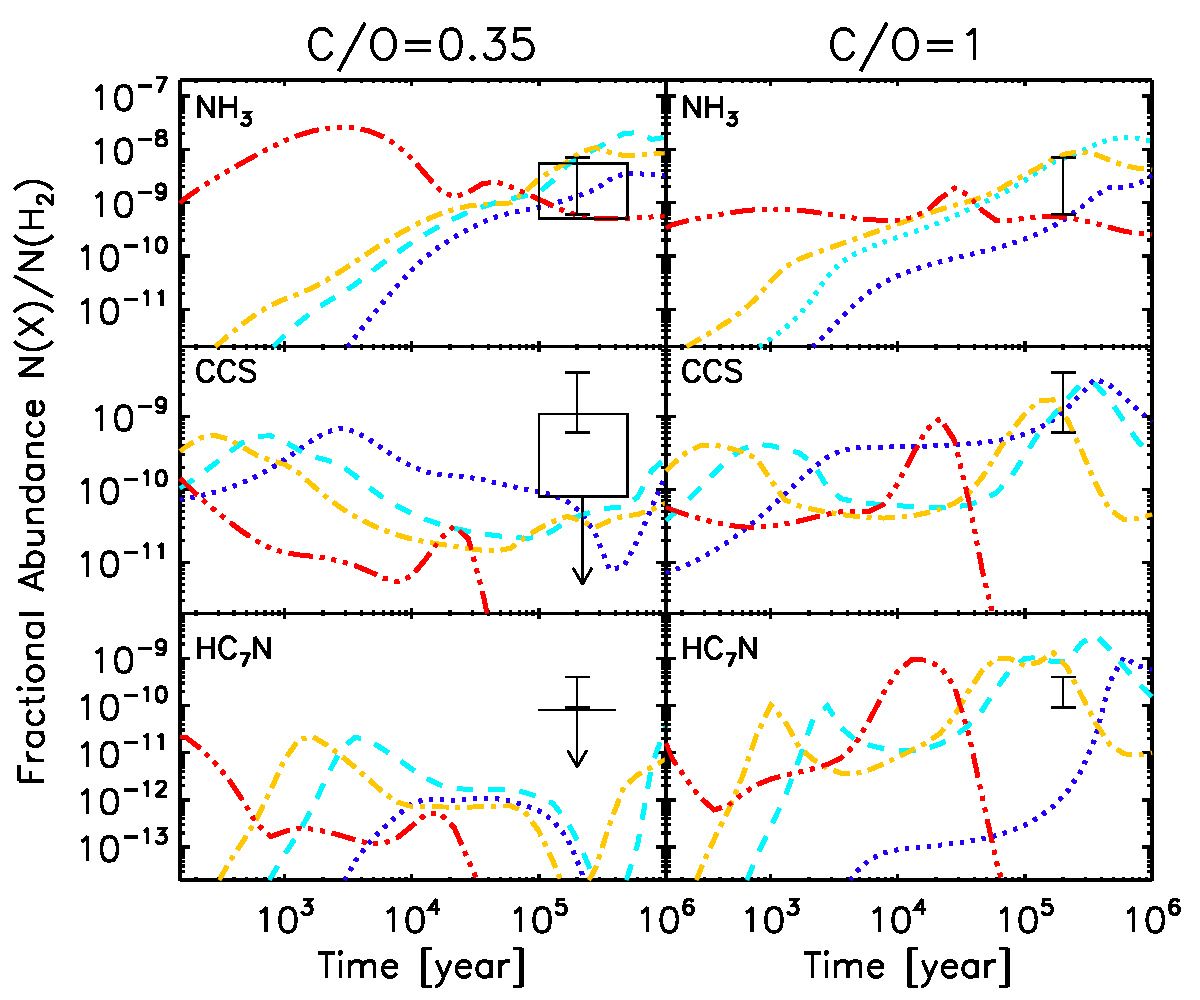}
\caption{{Fractional abundances of NH$_3$, CCS, and HC$_7$N with respect to H$_2$ as a function of time. The blue dotted, cyan dashed, orange dashed-dotted, and red dashed-triple dotted lines denote densities at n$_{\rm H_2}$ = 4 $\times$ 10$^3$, 2 $\times$ 10$^4$, 5 $\times$ 10$^4$, and 5 $\times$ 10$^5$ cm$^{-3}$, respectively. The chemical models with the initial gas phase C/O ratio of 0.35 (left column) have good agreement with most of the regions in L1495-B218 filaments except L1495A-N and L1521D. The chemical models with C/O ratio of 1 (right column) show relatively more-abundant CCS and HC$_7$N than the other models and better match the CCS and HC$_7$N abundances in L1495A-N and L1521D. The black boxes show ranges of the observed fractional abundances in starless cores expect L1495A-N and L1521D. The arrows denote non-detection of CCS and HC$_7$N in B10 and B213W. The bar plots denote ranges of the fractional abundances in L1495A-N and L1521D. }}
\label{chem}
\end{figure*}

We employ the following model parameters to represent the chemical evolution of CCS, NH$_3$ and HC$_7$N in the majority of starless cores in the L1495-B218 filaments.
\begin{itemize}
\item High-density core: T=8 K; n$_{\rm H_2}$=$5\times10^{5}$ cm$^{-3}$; A$_{\rm V}$=20 mag
\item Mid-density core: T=10 K; n$_{\rm H_2}$=$5\times10^{4}$ cm$^{-3}$; A$_{\rm V}$=15 mag
\item Low-density core: T=10 K; n$_{\rm H_2}$=$2\times10^{4}$ cm$^{-3}$; A$_{\rm V}$=8 mag
\item Core outskirts: T=15 K; n$_{\rm H_2}$=$4\times10^{3}$ cm$^{-3}$; A$_{\rm V}$=3 mag.
\end{itemize}
The densities are chosen from the average density along the lines-of-sight to the starless cores in the L1495-B218 filaments. The extinction values are directly estimated from the \citet{schmalzl10} observations. The temperature values are adopted from the kinetic temperature of the starless cores estimated using NH$_3$ \citep{seo15}.

The selection of the set of initial gas-phase elemental abundances in the chemical models is not a trivial task since the abundances of the elements strongly influence the chemical state of the gas \citep{agundez13}. A prime example is the relative amount of carbon and oxygen, which is known to strongly affect the chemistry \citep{vandishoeck98,lebourlot95}. To reflect the large uncertainties in the gas-phase elemental abundances of carbon and oxygen, the range of carbon-to-oxygen gas-phase elemental abundance ratio (noted C/O hereafter) is often varied between 0.3 and 1.5 \citep{cartledge04,legal14}. The elemental gas-phase abundance of sulfur is also very poorly constrained and we have yet to locate the main reservoir of sulfur in the ISM \citep{majumdar17,vidal18}. The range of initial sulfur abundances can vary by almost three orders of magnitude if we consider the range determined by the so-called low metal and high metal abundances defined by \citet{graedel82}.

Here, we firstly show chemical models with an initial C/O elemental ratio of 0.35, which is close to the solar value \citep{legal14}. The initial sulfur abundance in the models is 5$\times$10$^{-7}$ with respect to hydrogen nuclei whereas the solar value is 1.5 $\times$ 10$^{-5}$. Thus, we have applied a depletion factor (defined by the ratio of solar to adopted value in the model) of 30 from the gas phase which is smaller than the best-fit depletion factor of 200 adopted to reproduce the observations of different sulfur bearing species in L1544 pre-stellar core by \citet{vastel18}. They also discussed the variation of sulfur depletion for different sources. We find that the initial sulfur abundance of 5$\times$10$^{-7}$ results in the best-fits to the observed fractional abundances in the L1495-B218 filaments.

Figure \ref{chem} shows the fractional abundance profiles of NH$_3$, CCS, and HC$_7$N as a function of time. By running a grid of chemical models, we find that the chemical models with C/O = 0.35 provide the best-fit to our observations except for L1495A-N and L1521D since the models and observations have good agreement within a factor of five. We consider a factor of five as the reasonable uncertainty between the models and the observations since the uncertainties of the observed fractional abundances are typically a factor of three. The uncertainties of the NH$_3$ fractional abundance are mostly in the H$_2$ column density due to uncertainty in the dust properties, whereas the column densities of NH$_3$ have uncertainties typically lower than 30\% due to high signal-to-noise. The column densities of HC$_7$N and CCS depend on the excitation temperature. If the excitation temperature is 6 K, the column density will be 30\% higher than a case when T$_{\rm k}$ = T$_{\rm ex}$, while it can be three times higher if T$_{\rm ex}$ = 4 K. Since we have not observed other transitions of CCS and HC$_7$N, we could not measure the excitation temperature. A typical excitation temperature of CCS 2$_1$ $-$ 1$_0$ measured in Taurus is around 6 K \citep{suzuki92,wolkovitch97,vastel18}. The excitation temperature of HC$_7$N 21$-$20 is not measured in Taurus.

In this subsection, we will firstly discuss our chemical models with C/O = 0.35 and compare the models to our observations except the two dense cores, L1495A-N and L1521D observed to have converging-motion signatures. We will elaborate on these two dense cores in the following subsection since they show significantly different chemical characteristics compared to the other starless cores in the L1495-B218 filaments.

\begin{deluxetable} {| c |c | c | c | c | }
\tablewidth{0pt}
\tablecaption{The main production and destruction reactions with their relative contributions for NH$_3$, CCS and HC$_7$N at different ages for core outskirts (4$\times$10$^3$ cm$^{-3}$) models with initial C/O = 0.35. \label{reactions1}}
\tablehead{ Density  & Age  & Species  & Formation & Destruction \\
cm$^{-3}$ & yrs &  &  & }
\startdata
\hline
                &       &   {\bf NH$_3$}  &  NH$_4^+$ + e$^-$ $\rightarrow$ H + NH$_3$ (69.2\%)            & NH$_3$ + C$^+$ $\rightarrow$  H + HCNH$^+$ (62.9\%)   \\
                &       &                 &  H-ice + NH$_2$-ice $\rightarrow$ NH$_3$ (30.2\%)              & NH$_3$ + C$^+$ $\rightarrow$  C + NH$_3^+$ (31.5\%) \\
 \cline{3-5}
                & 10$^2$ & {\bf CCS}      &  HC$_2$S$^+$ + e$^-$ $\rightarrow$ H + CCS (89.2\%)            & CCS + C$^+$ $\rightarrow$  S$^+$ + C$_3$ (68.7\%)   \\
                &        &                & HC$_3$S$^+$ + e$^-$ $\rightarrow$ CH + CCS (8.4\%)             & CCS + C$^+$ $\rightarrow$  C + C$_2$S$^+$ (19.6\%) \\
 \cline{3-5}
                &        & {\bf HC$_7$N}  & N + C$_7$H$^-$ $\rightarrow$ HC$_7$N + e$^-$ (41\%)            & HC$_7$N + C$^+$ $\rightarrow$ H + C$_8$N$^+$ (49.7\%)  \\
                &        &                & CN + C$_6$H$_2$ $\rightarrow$ H+ HC$_7$N  (35.6\%)             & HC$_7$N + C$^+$ $\rightarrow$ CN + C$_7$H$^+$ (49.7\%)  \\
\cline{2-5}
                &        &   {\bf NH$_3$}  & H-ice + NH$_2$-ice $\rightarrow$ NH$_3$ (69.9\%)              & NH$_3$ + S$^+$ $\rightarrow$  S + NH$_3^+$ (34\%)   \\
                &        &                 &    NH$_4^+$ + e$^-$ $\rightarrow$ H + NH$_3$ (29.4\%)         & NH$_3$ + C$^+$ $\rightarrow$  H + HCNH$^+$ (20.5\%) \\
\cline{3-5}
4$\times$10$^3$ & 10$^5$ &  {\bf CCS}    &  HC$_2$S$^+$ + e$^-$ $\rightarrow$ H + CCS (50.4\%)            & CCS + O $\rightarrow$  CO + CS (61.6\%)   \\
                &        &               &  C + HCS $\rightarrow$ CCS + H (28.4\%)                        & CCS + C $\rightarrow$  C$_2$ + CS (30.3\%) \\
\cline{3-5}
                &        & {\bf HC$_7$N} & N + C$_8$H $\rightarrow$ HC$_7$N + C (85.4\%)                  & HC$_7$N + C $\rightarrow$ H + C$_8$N (47.3\%)  \\
                &        &               & C$_7$H$_2$N$^+$ + e$^-$ $\rightarrow$ H+ HC$_7$N  (8\%)        & HC$_7$N + H$^+$ $\rightarrow$ H + HC$_7$N$^+$ (16.1\%)  \\
\cline{2-5}
                &        &   {\bf NH$_3$} &  NH$_4^+$ + e$^-$ $\rightarrow$ H + NH$_3$ (96.5\%)           & NH$_3$ + H$^+$ $\rightarrow$  H + NH$_3$$^+$ (27.2\%)   \\
                &        &                &  H-ice + NH$_2$-ice $\rightarrow$ NH$_3$ (3.1\%)              & NH$_3$ + h$\nu$ $\rightarrow$  H + NH$_2$ (21.5\%) \\
\cline{3-5}
                & 10$^6$ &  {\bf CCS}     &  HC$_2$S$^+$ + e$^-$ $\rightarrow$ H + CCS (58\%)             & CCS + H$^+$ $\rightarrow$  H + C$_2$S$^+$ (61\%)   \\
                &        &                &  S + CCH $\rightarrow$ H + CCS (37.1\%)                       &  CCS + H$_3$$^+$ $\rightarrow$  H$_2$ + HC$_2$S$^+$ (10\%)  \\
\cline{3-5}
                &        & {\bf HC$_7$N}  &  CN + C$_6$H$_2$ $\rightarrow$ H+ HC$_7$N  (47.6\%)           & HC$_7$N + H$^+$ $\rightarrow$ H + HC$_7$N$^+$ (41.5\%)  \\
                &        &                &  C$_7$H$_2$N$^+$ + e$^-$ $\rightarrow$ H+ HC$_7$N  (40.4\%)   & HC$_7$N + H$_3$$^+$ $\rightarrow$ H$_2$ + C$_7$H$_2$N$^+$(49.7\%) \\
\enddata
\end{deluxetable}

\begin{deluxetable} { |c| c| c| c| c| }
\tablewidth{0pt}
\tablecaption{The main production and destruction reactions with their relative contributions for NH$_3$, CCS and HC$_7$N at different ages for mid-density core (5$\times$10$^4$ cm$^{-3}$) models with initial C/O = 0.35. \label{reactions2}}
\tablehead{ Density  & Age  & Species  & Formation & Destruction \\
cm$^{-3}$ & yrs &  &  & }
\startdata
\hline
                &        &   {\bf NH$_3$}   & H-ice + NH$_2$-ice $\rightarrow$ NH$_3$ (92\%)                 & NH$_3$ + C$^+$ $\rightarrow$  H + HCNH$^+$ (61.9\%)   \\
                &        &                  &   NH$_4^+$ + e$^-$ $\rightarrow$ H + NH$_3$ (7.9\%)            & NH$_3$ + C$^+$ $\rightarrow$  C + NH$_3^+$ (31\%) \\
 \cline{3-5}
                & 10$^2$ & {\bf CCS}        &  HC$_2$S$^+$ + e$^-$ $\rightarrow$ H + CCS (78.2\%)            & CCS + C$^+$ $\rightarrow$  S$^+$ + C$_3$ (65.7\%)   \\
                &        &                  &  HC$_3$S$^+$ + e$^-$ $\rightarrow$ CH + CCS (17.1\%)           & CCS + C$^+$ $\rightarrow$  C + C$_2$S$^+$ (18.8\%) \\
 \cline{3-5}
                &        & {\bf HC$_7$N}   & N + C$_7$H$^-$ $\rightarrow$ HC$_7$N + e$^-$ (56.8\%)          & HC$_7$N + C$^+$ $\rightarrow$ H + C$_8$N$^+$ (48.4\%)  \\
                &        &                 & CN + C$_6$H$_2$ $\rightarrow$ H+ HC$_7$N  (20.6\%)             & HC$_7$N + C$^+$ $\rightarrow$ CN + C$_7$H$^+$ (48.4\%)  \\
\cline{2-5}
                &        &  {\bf NH$_3$}   & NH$_4^+$ + e$^-$ $\rightarrow$ H + NH$_3$ (95.7\%)             & NH$_3$ + H$_3$$^+$ $\rightarrow$  H$_2$ + NH$_4^+$ (25.2\%)   \\
                &        &                 &     H-ice + NH$_2$-ice $\rightarrow$ NH$_3$ (3.3\%)            & NH$_3$ + H$_3$O$^+$ $\rightarrow$  H$_2$O + NH$_4^+$ (24.3\%) \\
\cline{3-5}
5$\times$10$^4$ & 10$^5$ &  {\bf CCS}      &  HC$_2$S$^+$ + e$^-$ $\rightarrow$ H + CCS (62.2\%)            & CCS + O  $\rightarrow$  CO + CS (90.2\%)   \\
                &        &                 &   HC$_3$S$^+$ + e$^-$ $\rightarrow$ CH + CCS (15.3\%)          & CCS + N $\rightarrow$  CN + CS (2.3\%) \\
\cline{3-5}
                &        &  {\bf HC$_7$N}  & N + C$_8$H $\rightarrow$ HC$_7$N + C (45\%)                    & HC$_7$N + H$_3$$^+$ $\rightarrow$ H$_2$ + C$_7$H$_2$N$^+$ (49.4\%)  \\
                &        &                 & C$_7$H$_2$N$^+$ + e$^-$ $\rightarrow$ H+ HC$_7$N  (44.5\%)     & HC$_7$N + HCO$^+$ $\rightarrow$ CO + C$_7$H$_2$N$^+$ (24.8\%)  \\
\cline{2-5}
                &        &  {\bf NH$_3$}   &  NH$_4^+$ + e$^-$ $\rightarrow$ H + NH$_3$ (99.5\%)            & NH$_3$ + H$_3$$^+$ $\rightarrow$  H$_2$ + NH$_4$$^+$ (76.2\%)   \\
                &        &                 &                                                                & NH$_3$ + H$^+$ $\rightarrow$  H + NH$_3^+$ (10.6\%) \\
 \cline{3-5}
                & 10$^6$ &  {\bf CCS}      &  HC$_2$S$^+$ + e$^-$ $\rightarrow$ H + CCS (73\%)              & CCS + H$_3$$^+$ $\rightarrow$  H$_2$ + HC$_2$S$^+$ (79.9\%)   \\
                &        &                 &  S + CCH $\rightarrow$ H + CCS (14.8\%)                        & CCS + H$^+$ $\rightarrow$  H + C$_2$S$^+$ (11.5\%) \\
 \cline{3-5}
                &        & {\bf HC$_7$N}   & C$_7$H$_2$N$^+$ + e$^-$ $\rightarrow$ H+ HC$_7$N  (53.1\%)     & HC$_7$N + H$_3^+$ $\rightarrow$ H$_2$ + C$_7$H$_2$N$^+$ (81.7\%)  \\
                &        &                 &   CN + C$_6$H$_2$ $\rightarrow$ H+ HC$_7$N  (45.5\%)           & HC$_7$N + H$^+$ $\rightarrow$ H + HC$_7$N$^+$(11.9\%) \\
\enddata
\end{deluxetable}

The general trend of the fractional abundances of NH$_3$, CCS, and HC$_7$N in the models with C/O = 0.35 show similar evolution to Aikawa's models with small differences in details. The main chemical channels for forming and destroying NH$_3$, CCS, and HC$_7$N are given in Table \ref{reactions1} and Table \ref{reactions2} for the core outskirts and mid-density core models. The fractional abundance of NH$_3$ increases almost monotonically as a function of time except for the high-density core model (see Figure \ref{chem}). In the high-density model, the NH$_3$ fractional abundance decreases after 3$\times$10$^3$ yrs due to depletion on to dust grains. The peak NH$_3$ abundance is lower when the gas density is higher since the depletion of NH$_3$ onto the dust grain surface becomes more efficient at the high-density core model, which is also observed in Aikawa's model. The fractional abundance of CCS shows a more complicated evolution compared to NH$_3$ with three different trends as a function of time: a rapid increase of the fractional abundance at early times ($<$a few 1000 yrs), a decrease of the fractional abundance from $\sim$10$^{-9}$ to 10$^{-11}$ from $\sim$1000 yrs to $\sim$0.1 Myr, and a slow increase of the abundance again after $\sim$0.1 Myr. The decrease of the CCS abundance is mainly due to its destruction by oxygen in gas-phase and is not due to depletion on dust grains (Table \ref{reactions1} and Table \ref{reactions2}), unlike other carbon-bearing species \citep[e.g.,][]{tafalla04,pagani05}. The slow increase in CCS after $\sim$0.1 Myr is because its destruction rate by oxygen is significantly reduced due to the depletion of oxygen on dust grains. At this stage, the formation rate of CCS is still relatively low so the CCS abundance does not rapidly increase as it does at earlier times. HC$_7$N also shows chemical trends similar to CCS except at low density where the HC$_7$N formation via N + C$_8$H $\rightarrow$ C + HC$_7$N is much slower than CCS.

We compared our observations to the chemical models with the initial gas-phase C/O = 0.35 in Figure \ref{chem}. We found that most regions, including B10, B213W, B211, B216, and B218 show good agreement with the models. For example, young and relatively less-condensed starless cores in B211 and B216,  have NH$_3$ fractional abundance $\sim$1 $\times$ 10$^{-9}$ with respect to H$_2$ at the core centers and have CCS fractional abundances of 0.4 $-$ 1 $\times$ 10$^{-9}$ with respect to H$_2$ at CCS peak-intensity positions. Considering the uncertainties in the fractional abundances, we found that B211 and B216 have better agreement with the core outskirts model. From this model, fractional abundances of NH$_3$ and CCS are 2.2 $\times$ 10$^{-9}$ and 3 $\times$ 10$^{-10}$, respectively, with respect to H$_2$ at a chemical age of $\sim$0.1 Myr which is reasonable for young starless cores. The HC$_7$N abundance at this chemical age is 1 $\times$ 10$^{-12}$ with respect to H$_2$ which explains our non-detection as well. The NH$_3$ and CCS abundances in B211 are well-reproduced by our models. For B216, NH$_3$ is in good agreements with the model at 0.1 Myr but CCS is a factor of five higher compared to the models at the same age.

Starless cores in B10 and B213W, which are more-condensed than the ones in B211 and B216, have fractional abundances of NH$_3$ of 3 $\times$ 10$^{-9}$ and 5 $\times$ 10$^{-9}$, respectively, at their centers and 1 $\times$ 10$^{-9}$ and 2 $\times$ 10$^{-9}$, respectively, at their outskirts. Considering the uncertainties in the fractional abundances, we think that the these two regions may correspond to a chemical age around 0.3 Myr or older, where the CCS fractional abundance steeply decreases in core outskirts models. At this age, the NH$_3$ abundance is 1 $-$ 20 $\times$ 10$^{-9}$ depending on density, whereas the CCS abundance is 0.1 $-$ 2 $\times$ 10$^{-10}$ with respect to H$_2$, which mostly falls below the lowest detected fractional abundance of CCS of 4 $\times$ 10$^{-10}$. We did not detect HC$_7$N in B10 and B213W and the fractional abundance of HC$_7$N in the model is $\sim$10$^{-12}$ or lower which is below the lowest observed fractional abundance of 5 $\times$ 10$^{-11}$ in our entire HC$_7$N survey.

Starless cores in B218 are more-condensed cores similar to those in B10 and B213W. These starless cores show weak CCS rings while we did not detect any HC$_7$N emission at our noise level. The NH$_3$ fractional abundance ranges 3$-$4$\times$10$^{-9}$ at the core centers. The NH$_3$ and CCS fractional abundances are 0.5$-$1$\times$10$^{-9}$ and 3$-$6 $\times$ 10$^{-10}$, respectively, at the outskirts. Comparing these abundances to the mid-density and outskirts models with the initial C/O = 0.35, we found that these starless cores may be at a chemical age between 0.1 Myr and 0.2 Myr. At this age, the mid-density and outskirts models with the initial C/O = 0.35 show the NH$_3$ and CCS fractional abundance ranges of 0.8$-$10 $\times$ 10$^{-9}$ and 0.5$-$1 $\times$ 10$^{-10}$ respectively and are in good agreements with the observed abundances.

We found that most of the regions in the L1495-B218 filaments agree with the chemical models with C/O = 0.35. However, we found that two starless cores, L1495A-N and L1521D, have significantly different characteristics compared to other starless cores in Taurus. In following section, we elaborate chemical evolution of L1495A-N and L1521D and the implications related to their dynamics.

\subsubsection{Chemistry of L1495A-N and L1521D}

\begin{deluxetable} { |c | c | c | c | }
\tablewidth{0pt}
\tablecaption{The main production and destruction reactions with their relative contributions for NH$_3$, CCS and HC$_7$N at the age between 0.2 and 0.4 Myr for core-outskirts and mid-density core models with initial C/O = 1. \label{reactions3}}
\tablehead{ Density    & Species  & Formation & Destruction \\
cm$^{-3}$ &  &    & }
\startdata
\hline
                  & {\bf NH$_3$} & H-ice + NH$_2$-ice $\rightarrow$ NH$_3$ (90.4\%)        & NH$_3$ + C$^+$ $\rightarrow$  H + HCNH$^+$ (40.2\%)   \\
                  &              &   NH$_4^+$ + e$^-$ $\rightarrow$ H + NH$_3$ (9.5\%)     & NH$_3$ + C$^+$ $\rightarrow$  C + NH$_3^+$ (20.1\%) \\
\cline{2-4}
4$\times$10$^3$   &  {\bf CCS}   &  HC$_2$S$^+$ + e$^-$ $\rightarrow$ H + CCS (68.9\%)     & CCS + C$^+$ $\rightarrow$  S$^+$ + C$_3$ (41.4\%)   \\
                  &              &  HC$_3$S$^+$ + e$^-$ $\rightarrow$ CH + CCS (27.1\%)    & CCS + C $\rightarrow$  C$_2$ + CS (25.8\%) \\
\cline{2-4}
                  & {\bf HC$_7$N}& N + C$_8$H $\rightarrow$ C+ HC$_7$N (68.6\%)            & HC$_7$N + C$^+$ $\rightarrow$ H + C$_8$N$^+$ (32.9\%)  \\
                  &              & CN + C$_6$H$_2$ $\rightarrow$ H+ HC$_7$N  (22\%)        & HC$_7$N + C$^+$ $\rightarrow$ CN + C$_7$H$^+$ (32.9\%)  \\
\hline
                  & {\bf NH$_3$} & NH$_4^+$ + e$^-$ $\rightarrow$ H + NH$_3$ (98.6\%)      & NH$_3$ + H$_3$$^+$ $\rightarrow$  H$_2$ + NH$_4^+$  (63.7\%)   \\
                  &              &                                                         & NH$_3$ + HCO$^+$ $\rightarrow$  CO + NH$_4^+$ (7\%) \\
\cline{2-4}
5$\times$10$^4$   &  {\bf CCS}   &  HC$_2$S$^+$ + e$^-$ $\rightarrow$ H + CCS (63\%)       & CCS + H$_3$$^+$ $\rightarrow$  H$_2$ + HC$_2$S$^+$ (71.2\%)   \\
                  &              & S + CCH $\rightarrow$ H + CCS (20.2\%)                  & CCS + H$^+$ $\rightarrow$  H + C$_2$S$^+$ (7.6\%) \\
\cline{2-4}
                  & {\bf HC$_7$N}&  C$_7$H$_2$N$^+$ + e$^-$ $\rightarrow$ H+ HC$_7$N  (50.9\%)  & HC$_7$N + H$_3$$^+$ $\rightarrow$ H$_2$ + C$_7$H$_2$N$^+$ (76.7\%)  \\
                  &              & CN + C$_6$H$_2$ $\rightarrow$ H+ HC$_7$N  (44.7\%)           & HC$_7$N + H$^+$ $\rightarrow$ H + HC$_7$N$^+$ (8.2\%)  \\
\enddata
\end{deluxetable}

The physical conditions of L1495A-N and L1521D including density structures, extinction at the core surface, kinetic temperature, and turbulence are similar to those of the starless cores in B213W and B10, but the fractional abundances of CCS and HC$_7$N in L1495A-N and L1521D are much higher than in the starless cores in B213W and B10.

The fractional abundances of NH$_3$ and CCS with respect to H$_2$ at the core centers (NH$_3$ peak-intensity position) are 4 $\times$ 10$^{-9}$ and 6 $\times$ 10$^{-10}$, respectively, for L1495A-N, and 5 $\times$ 10$^{-9}$ and 3 $\times$ 10$^{-10}$, respectively, for L1521D. We did not detect any HC$_7$N at the core centers. At the CCS peak-intensity position at the outskirts of the cores, the fractional abundances of NH$_3$ and CCS with respect to H$_2$ are 3 $\times$ 10$^{-10}$ and 3 $\times$ 10$^{-9}$ respectively for L1495A-N, and 5 $\times$ 10$^{-10}$ and 2 $\times$ 10$^{-9}$, respectively, for L1521D. We did not detect HC$_7$N emission at these positions. On the other hand, at the HC$_7$N peak-intensity position, the fractional abundances of NH$_3$, CCS, and HC$_7$N with respect to H$_2$ are 6 $\times$ 10$^{-10}$, 1 $\times$ 10$^{-9}$, and 2 $\times$10 $^{-10}$, respectively, for L1495A-N and 2 $\times$ 10$^{-9}$, 1 $\times$ 10$^{-9}$, and 3 $\times$ 10$^{-10}$, respectively, for L1521D.

For L1495A-N and L1521D, we found that the models with n$_{\rm H_2}$ = 4 $\times$ 10$^{3}$ cm$^{-3}$ and 5 $\times$ 10$^4$ cm$^{-3}$ correspond to the average densities in the core outskirts and the core center, respectively, evaluated by column density/size. The actual volume densities of the core centers may be several times higher than 5$\times$10$^4$ cm$^{-3}$ but still less than 5 $\times$ 10$^5$ cm$^{-3}$ if their density profiles are similar to the Bonnor-Ebert sphere \citep{ebert55,bonnor56}. Comparing the observed NH$_3$ and CCS abundances to the models with the initial C/O = 0.35, we found that the observed CCS fractional abundances of L1495A-N and L1521D are significantly higher (at least 20 times more abundant at core center and at the CCS peak-intensity position) than the models with the initial C/O = 0.35 if we determine the age of the cores using the observed NH$_3$ abundance to be $\sim$0.1 Myr. Also, the peak CCS abundances in the models are still more than 5 times lower compared to the observed CCS fractional abundance at the CCS peak-intensity positions of L1495A-N and L1521D. We found that the observed HC$_7$N fractional abundances at the HC$_7$N peak-intensity position are at least 100 times higher than the fractional abundance predicted by the model with the initial C/O = 0.35. These discrepancies between the observed fractional abundances and the models with the initial C/O = 0.35 strongly suggest that the two cores are going through a different chemical evolutionary path compared to the other starless cores in the L1495-B218 filaments.

Table \ref{reactions1} and Table \ref{reactions2} clearly show the importance of the O + CCS$\rightarrow$ CO + CS reaction that controls the chemistry of CCS in the majority of starless cores in the L1495-B218 filaments described in earlier subsection. By running a grid of chemical models, we found that the reaction O + CCS$\rightarrow$ CO + CS is highly sensitive to the initial gas phase C/O ratio since the abundance of the gas phase atomic oxygen controls the reaction flux of this channel i.e. the higher the C/O ratio, the lower the rate of destruction of CCS. To explain the observed high abundances of CCS together with HC$_7$N within a reasonable uncertainty, we have found the best agreement to be with the chemical models having C/O=1.

The right column of Figure \ref{chem} shows the evolution of the fractional abundances as a function of time. The general trend of the NH$_3$ fractional abundance over time in the C/O = 1 models is similar to that of the C/O = 0.35 models. On the other hand, the evolution of the CCS and HC$_7$N fractional abundances in the C/O = 1 models is very different from the C/O = 0.35 models. The CCS and HC$_7$N fractional abundances in the C/O = 1 models are significant even at an age of $>$1 Myr, having peak abundances $>$10$^{-9}$, while the CCS and HC$_7$N fractional abundances in the C/O = 0.35 models tend to decrease and maintain an abundance of $<$10$^{-10}$ over a long period of time after a few times 0.01 Myr. This is mainly due to less oxygen being available to destroy CCS and more carbon to form CCS and HC$_7$N when the initial gas-phase C/O ratio is high.

The observed fractional abundances of NH$_3$ and CCS at the center of L1495A-N and L1521D correspond to chemical ages between 0.2 and 0.4 Myr using the mid-density model which gives the modeled abundances of 4$\times$10$^{-9}$ for NH$_3$ and 2$\times$10$^{-10}$ for CCS. Table \ref{reactions3} shows the major reaction channels for production and destruction of NH$_3$, CCS and HC$_7$N at the age betweeen 0.2 and 0.4 Myr for core outskirts and mid-density core model. If we use the high-density model, we have a chemical age matching the observed NH$_3$ abundance arrund 0.01 Myr. The high-density model results in younger ages due to the low abundance of CCS after 0.03 Myr. Considering that we have average densities of 0.5$-$2 $\times$ 10$^5$ cm$^{-3}$ toward the core centers of L1495A-N and L1521D, we do not think that the two cores have central densities near to 5 $\times$ 10$^5$ cm$^{-3}$. Also, we do not observe any central depletion of NH$_3$ in L1495A-N and very little depletion of NH$_3$ in L1521D (only 20\% decrease of the NH$_3$ fractional abundance), while the high-density model shows an order of magnitude decrease of the NH$_3$ fractional abundance. The density is also unlikely to reach the central density of 5 $\times$ 10$^5$ cm$^{-3}$ within $<$0.01 Myr since a dense core typically takes much longer than 0.01 Myrs to reach central density 5 $\times$ 10$^5$ cm$^{-3}$ \cite[e.g.,][]{gong09,gong11}. Thus, chemical ages of L1495A-N and L1521D using the mid-density core model, 0.2$-$0.4 Myr, fit better to the observed quantities and are similar to the ages of other dense cores \citep[e.g.,][]{ward-thompson07,brunken14}.

The observed fractional abundances of NH$_3$ and CCS at the outskirts/CCS peak-intensity position of L1495A-N and L1521D also show good agreement with the outskirts model at an age between 0.2 and 0.4 Myr, which gives NH$_3$ and CCS fractional abundances 6$\times$10$^{-10}$ and 2$\times$10$^{-9}$ respectively. This age range is relatively consistent with the chemical age at the core centers of L1495A-N and L1521D and also similar to the chemical ages of the other starless cores in the L1495-B218 filaments 0.3 Myr or older.

The comparisons between observations and the chemical models suggest that the majority regions of the L1495-B218 filaments have likely formed in an environment with the initial C/O ratio in the gas around 0.35, which is close to the solar value \citep{legal14}. However, for the two starless cores L1495A-N and L1521D, we were not able to explain high abundance of CCS and HC$_7$N unless we adopt chemical models with higher initial gas phase C/O ratio of $\geq$1. Based on the fact that most of the cores in the L1495-B218 filaments agrees with the chemical model with a low initial C/O, the initial C/O ratio across the Taurus molecular cloud was likely homogeneous with a value close to solar. Therefore, it appears likely that L1495A-N and L1521D underwent different formation mechanism or different evolutionary path compared to the rest of the cores in the L1495-B218 filaments.

While the physics of having a high initial gas-phase C/O ratio is out of the scope of this current paper, we suggest two possible mechanisms that could produce results similar to the chemical model with a high initial C/O ratio in the gas \citep[e.g.,][ and reference therein]{hincelin11}. The first possible mechanism is that a core forms by a collision of filaments. Generally, the gas-phase C/O ratio varies significantly depending on the gas density, extinction, and age. At a molecular cloud clump density $>$1000 cm$^{-3}$, a gas-phase C/O ratio can increase significantly after $\sim$1 Myr due to oxygen depletion. Thus, if molecular filaments are older than $\sim$1 Myr and the collision of the filaments triggers core formation, the resulting cores would form in a high C/O ratio environment and have bright CCS and HC$_7$N emission. The second possible mechanism is that a core forms in a low C/O ratio environment (similar to the solar value), spends a long time in a quasi-static stage, and collapses while accreting gas from surrounding filament materials. The aged filament gas would have a high C/O ratio in the gas, so the accreted gas on the surface of the core may develop abundant CCS and HC$_7$N and can make bright CCS and HC$_7$N rings.

\subsection{A Stellar Association/Cluster Forming Candidate: L1495A/B7N}

L1495A/B7N is the hub and the most evolved region of the L1495-B218 filaments. Seventeen out of 35 young stellar objects along the filaments are associated with L1495A/B7N. L1495A/B7N has the highest column density along the filament and is also the most massive filament. Our NH$_3$ survey shows that L1495A/B7N is a gravitationally unstable clump with two NH$_3$ intensity peaks in L1495A-S and two NH$_3$ intensity peaks in L1495A-N \citep{seo15}. CCS and HC$_7$N in L1495A-N are as bright as NH$_3$ due to recently condensed gas but the intensity peaks are displaced from the NH$_3$ peak. HCN \& HCO$^+$ 1$-$0 observations show strong blue asymmetric, self-absorbed profiles, which indicate converging-motions. In addition, gas flows in $^{12}$CO are connected with flows seen in CCS and NH$_3$ in the Position-Position-Velocity space. L1495A-N is currently gravitationally accreting gas and is also being fed gas by a large-scale flow.

We compare L1495A-N with theoretical models of quasi-static collapse of dense cores \citep{larson69,penston69,foster93,gong09,seo13} to have an idea whether the inward flows toward L1495A-N are mainly due to gravitational collapse or to large-scale colliding flows. The theoretical models evaluate the minimum and the maximum converging-motion speeds due to gravitational collapse for a given density concentration. We compare the converging flow speed and the dimensionless dense core radius $\xi_{\rm max}$ of L1495A-N with those of the theoretical models. The $\xi_{\rm max}$ of L1495A-N is approximately 8, which is estimated by taking the long axis of L1495A-N measured in \citet{seo15} and the mean density calculated by dividing the column density by the size of the long axis. The calculation shows that L1495A-N has a converging flow roughly seven times faster than the maximum converging-motion speed that the gravitational collapse of a dense core can have at $\xi_{\rm max}$ of 8. The converging flow speed of L1495A-N even exceeds the maximum converging-motion speed of gravitational collapse during the whole evolution that are predicted by the theoretical models ($<$Mach 4, \citealt{larson69,penston69,foster93}). Also, the position of the fastest converging motion in L1495A-N does not match with the fastest infall layer in the theoretical models, which is very near to the core center in the last stage of collapse. Thus, L1495A-N is collapsing with an exceptional speed and likely has its dynamics driven by large-scale inflows.

We estimate the star formation rate in L1495A-N considering gas fed by large-scale inflows. We use the inflow velocity of 0.8 km s$^{-1}$ measured in HCO$^+$ 1$-$0. We take the lowest column density along the 5$\sigma$ SNR edge of CCS emission in our map (N = 5 $\times$ 10$^{21}$ cm$^{-2}$) and the long-axis of L1495A-N seen in CCS (0.7 pc). We assume a star formation efficiency of 30\%, which is the average efficiency for dense structures in molecular clouds \citep{andre14}. Our estimate gives 16 M$_\odot$/Myr. L1495A/B7N is already associated with multiple young stellar objects and is likely to keep making stars, forming a stellar association/cluster, unless large-scale inflows are disturbed by radiation pressure or outflows from protostars.

\subsection{Three Modes of Star formation in the L1495-B218 filaments and Implications to Low-Mass Star Formation in Molecular Clouds}

\begin{figure*}[tb]
\centering
\includegraphics[angle=0,scale=0.33]{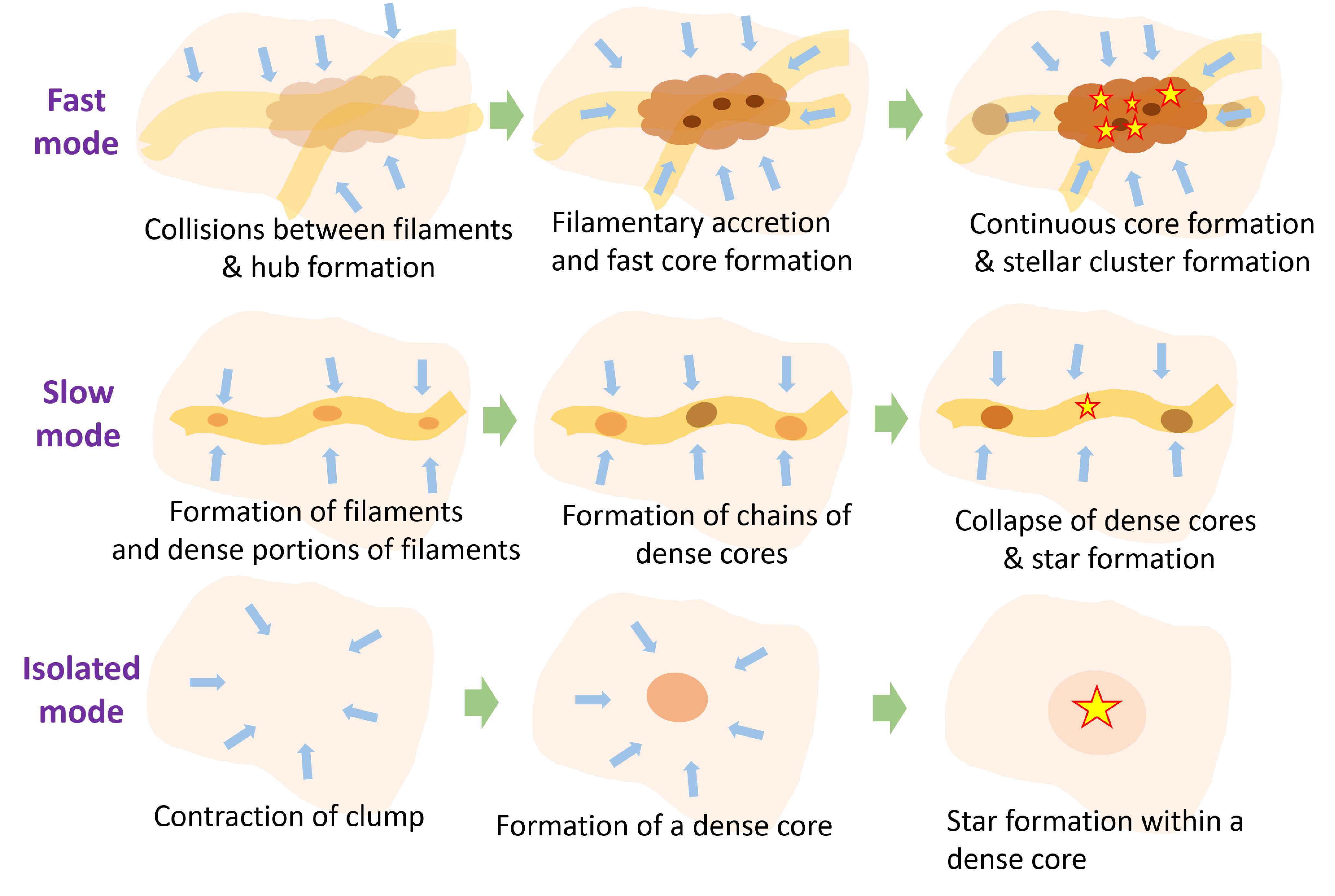}
\caption{{ The three modes of star formation. The first (fast) mode describes star formation at the hub, where the column density is highest and the global gravitational potential well is deepest. Large-scale flows drive continuous star formation in relatively short time and form a stellar cluster/association. The second (slow) mode elaborates the formation of the dense core chains within filaments due to gravitational fragmentation of filaments and localized star formation within each core. The third (isolated) mode shows formation of an isolated dense core and localized star formation removed from filamentary structures, which is often discussed as conventional star formation(e.g, L1544, Bok globules).  }}
\label{three_SF}
\end{figure*}

An important question in star formation within molecular cloud complexes is how the dynamical and chemical evolution in a cloud-scale structure are connected with and affect the small-scale dense structure in star formation. Through our new CCS, HC$_7$N, HCN, and HCO$^+$ observations along with observations in NH$_3$, CO isotopes, and far-IR, we think that there may be three different star formation processes taking place in the L1495-B218 filaments (Figure \ref{three_SF}): stellar cluster/association formation at the hub of filaments (fast mode), star formation in a chain of dense cores within a filament (slow mode), and star formation in an isolated dense core (isolated mode).

Analysis of the gas motions in the hub (L1495A/B7N) suggests that large-scale dynamics are playing a considerable role in not only forming dense cores but also collapsing dense cores and forming stars. The hub is the densest and the most massive part of the filaments and likely formed through a collision of two or three filamentary structures. Observations in HCN, HCO$^+$ and $^{13}$CO show that a large-scale converging flow (inferred from the blue asymmetry in HCN and HCO$^+$ 1$-$0 and the velocity gradients along the three groups of filaments) is still feeding the hub and driving collapse of cores. This ultimately promotes star formation since accreted gas adds more mass to dense cores, makes cores unstable, drives them to collapse, and also keeps forming new dense cores. Also, the ram pressure of a converging flow compresses dense cores to form protostars in a relatively short timescale \citep{gong09,seo13}. Thus, in the hub, stars are expected to form fast and continuously unless the large-scale flows are disturbed, or the surrounding gas is depleted.

A chain of dense cores embedded within a filament (e.g., L1521D, No.30, No.31, and No.32 in B213E, which are not at the hub) is under influence of surrounding filaments and neighboring dense cores but their collapse may not be strongly related to their surroundings. The virial analysis of NH$_3$ dense cores within filaments shows that only some of the dense cores within filaments are gravitationally bound, while a number of the dense cores are pressure-confined within sterile filaments by the weight of the filament material surrounding dense cores and the ram pressure of a large-scale converging flow \citep{seo15}. Recent NH$_3$ surveys toward Orion A and Cepheus have also reported that there are a number of pressure-confined cores \citep{keown17, kirk17}. The denser portions of filaments likely grew as dense cores as the filaments gained more mass through accreting ambient gas or converging flows of turbulent cells \citep{kirk13, palmeirim13, dhabal18}. Thus, the formation of dense cores is closely related with the formation and evolutionary history of the filaments. However, the subsequent collapse of dense cores does not seem to be directly related to the dynamics of filaments nor to a large-scale flow in CO as we show in L1521D. In this star formation process, the large-scale dynamics may have played a considerable role in forming filaments and dense cores but not in the final stage of forming a star within a dense core, which is similar to the {\it fray and fragment} model \citep{tafalla15} except that we suggest a formation mechanism of the pressure-confined cores in sterile filaments in addition to the core formation through gravitational fragmentation of the molecular filaments \cite[e.g.,][]{schmalzl10,hacar11,tafalla15}.

Finally, we found only one isolated dense core, core No.17, in the L1495-B218 filaments \citep{seo15}. It is not associated with any of the velocity-coherent filaments identified by \cite{hacar13} and also does not connect to the three groups of the filaments shown in Figure \ref{l1495a_channel_13co}. Since it is detached from the filaments and its surroundings are ambient gas, it is not likely influenced by filaments, nor the large-scale dynamics of the filaments. Virial analysis of the core shows that it is in a marginally critical state and that its profile resembles the Bonnor-Ebert sphere, which are similar to characteristics found in other isolated cores and Bok globules \citep[e.g.][]{bacmann00, kandori05}. This suggests that the core No.17 is evolving in a quasi-static manner in a relatively quiescent environment. The isolated mode has been studied using the brightest cores (e.g., L1544, Bok globules, etc.), however, the recent Herschel surveys showed that that majority of dense cores are embedded within filaments (e.g., $>$60\% in Aquila-Rift, \citealt{andre10}). Thus, the isolated mode is not likely a dominant mode of star formation in molecular clouds.

The three modes of star formation may also be found in other low-mass star-forming clouds. One of the examples with deep kinematic studies is Serpens-South. L1495A/B7N and the hub of Serpens-South share several similar characteristics such as bright carbon-chain molecules including CCS and cyanopolyynes along with bright NH$_3$ emission, an association of a stellar cluster, the signatures of colliding flows, and large-scale accretion toward the hub \citep{kirk13, friesen13, nakamura14}. The velocity gradients of large-scale flows seen in $^{12}$CO seem to be also similar between the two regions ($\sim$2 km s$^{-1}$ pc$^{-1}$) \citep{goldsmith08,tanaka13}. These characteristics strongly suggest that the hub of Serpens-South is also in the fast mode of star formation but in a scaled-up version compared to L1495A/B7N (L1495A/B7N: 0.7 pc, 60 M$_\odot$, 17 YSOs, \citealt{seo15}; Serpens-South: 10 pc, 360 M$_\odot$, 90 YSOs, \citealt{kirk13}). Within the filamentary regions of Serpens-South, there are small numbers of protostars with the fraction to the protostar population being much lower compared to the one at the hub, which is the same characteristic to the slow star formation mode in the L1495-B218 filaments in Taurus. Other than Serpens South, the hub-filament structures in star-forming regions are very common across low-mass to high-mass star-forming regions \citep{myers09} and the fast mode or a similar star formation mode is often found be the dominant processes in those regions (e.g., Serpens: \citealt{duarte-cabral10,duarte-cabral11}, OMC1-A: \citealt{hacar17}, G22: \citealt{yuan18}). Our study demonstrates that the Taurus molecular cloud, which is relatively less active in star formation compared to Orion, IRDC, Serpens, etc., also has a similar mode of star formation, and might indicate an existence of a scalable star formation mode across star-forming regions.

The three modes of low-mass star formation may result in different initial-mass-functions of stars (IMF). The IMF is one of the most important properties of star formation because it determines the evolution of stellar clusters and galaxies, and it has been extensively studied in both observation and theory to find realistic IMFs and to understand where it originates from. The clump-mass-function (CMF) is often studied to understand the origin of IMF \citep[e.g.][]{offner14} and some studies reported that the CMF and the IMF share a similar form. However, based on our study, the CMF may not be directly related to the origin of the IMF. The main contributors to the CMF are the chains of dense cores within the filaments. In the L1495-B218 filaments, 34 out of 39 NH$_3$ cores are within the filaments, while we found only four NH$_3$ cores at the hub and one isolated dense core. For the IMF, on the other hand, the main contributors are protostars at the hub because half of the protostars in the L1495-B218 filaments are associated with the hub, and more protostars are expected to form in the hub. This trend is also found in other molecular clouds. \citet{myers13} showed multiple examples of hub structures in nearby molecular clouds and IRDCs and demonstrated that these hub structures are associated with stellar groups or clusters. On the other hand, dense cores are typically not resolved due to the complexity of the hubs and observation limitations. It will be interesting to study how the different star formation modes affect the IMF.

\section{SUMMARY}

In this paper, we have presented CCS and HC$_7$N maps of the L1495-B218 filaments in the Taurus molecular cloud with an angular resolution of 31$''$ and a velocity resolution of 0.038 km s$^{-1}$. Using the maps, we investigated the chemistry and kinematics of the star-forming regions L1495A/B7N and B213E by comparing CCS and HC$_7$N to NH$_3$, 500 $\mu$m dust continuum, $^{12}$CO, and C$^{18}$O. The main results are as follows:

1. We carried out a complete survey in CCS and HC$_7$N along the L1495-B218 filament at noise levels of 120 mK for HC$_7$N $J$ = 21 $-$ 20 and 154 mK for CCS $J_N$ = 2$_1$ $-$ 1$_0$. We found strong CCS emission ($>$1 K) toward two more evolved regions, L1495A/B7N and B213E, and one less evolved region, B216. We observed weak CCS emission ($<$1 K) in a more evolved region, B218, and a less evolved region, B211. We found no detectable CCS emission in B10 and B213W at our noise level. Our survey shows that CCS emission is not related to the global evolutionary stages of regions.

2. CCS intensity peaks are not coincident with NH$_3$ and dust intensity peaks in most of the dense cores. Only in NH$_3$ core No.32 CCS, NH$_3$, and 500$\mu$m dust continuum intensity peaks coincide in position. CCS emission in more evolved regions, L1495A/B7N and B213E, shows an arc-like or a ring-like shape. L1521B (in B216), which was previously reported to be a young CCS-bright core also does not have CCS emission that coincides with the NH$_3$ or dust continuum emission peaks. These examples show that CCS rarely traces core structure.

3. We found through a grid of chemical models that the fractional abundance of CCS and HC$_7$N are significantly sensitive to the initial C/O ratio in the gas. In the model with the initial gas phase C/O ratio being 0.35, which is close to the solar value, the evolution of the fractional abundances of CCS and HC$_7$N is similar to the chemical evolution shown in \citet{suzuki92} and \citet{aikawa01}, where CCS and HC$_7$N are relatively abundant at a young age ($<$0.1 Myr) but get destroyed later times ($>$0.5 Myr). On the other hand, when the initial C/O ratio in the gas is high ($\geq$1), CCS and HC$_7$N are abundant for a long time. This suggests that CCS and HC$_7$N may be good tracers to constrain the initial C/O ratio in the gas rather than the evolutionary stage of dense cores. A future survey including multiple star-forming regions is required to confirm a correlation between the CCS and HC$_7$N fractional abundances and the initial gas-phase C/O ratio.

4. The velocity dispersions of CCS lines range from 0.15 km s$^{-1}$ to 0.3 km s$^{-1}$ with the majority of CCS spectra (94\%) having velocity dispersion less than the average sound speed of the L1495-B213 filaments (0.19 km s$^{-1}$). Considering the large molecular weight of CCS and relatively small thermal velocity (0.04 km s$^{-1}$ at 10 K), this suggests that nonthermal motions in CCS bright regions are subsonic, which is the same conclusion we obtained from our NH$_3$ survey. The transition-to-coherence is not observed at the sensitivity level of our observations.

5. L1495A/B7N is likely a stellar association/cluster forming candidate. From the kinematics and chemistry of NH$_3$, CCS, $^{12}$CO, HCN, and HCO$^+$, we find that a supersonic converging flow toward L1495A-N and that the flow is connected to a large-scale flow ($>$1 pc) seen in $^{13}$CO. L1495A/B7N is already associated with multiple protostars and is likely to form more protostars due to the converging flow. This may lead to formation of a stellar association or cluster in the region.

6. L1521D is a slowly contracting dense core. The converging-motion is not connected with a large scale flow seen in $^{12}$CO. Its converging-motion speed is subsonic, which is consistent with a spontaneous collapse model due to self-gravity.

Finally, comparing the star formation processes in L1495A-N, L1521D, and isolated dense cores, we conclude that there are three different star formation modes in the L1495-B218 filaments. The first (fast) mode is star formation at the hub driven by a large-scale flow (e.g., L1495A/B7N) from the formation of filaments and dense cores to the collapse of dense cores. The second (slow) mode is the slow collapse of dense cores within filaments (e.g., L1521D). The third (isolated) mode is star formation through the quasi-static collapse of isolated starless cores (e.g., core No.17). Since other low-mass star-forming clouds share similar characteristics (hub-filament and isolated dense cores), the three processes are likely to be found in other molecular clouds as well. It will be important to study how each mode affects the evolution of molecular clouds and IMF of protostars and which is dominant process in other molecular clouds.

\acknowledgments
We thank the anonymous referee for providing constructive comments that have improved the contents of this paper. This research was carried out in part at the Jet Propulsion Laboratory, California Institute of Technology, under a contract with the National Aeronautics and Space Administration. We thank Hugo Medrano for his contributions to Argus. YS and YLS were partially supported by NSF grant AST-1410190. YS and LM acknowledges support from the NASA postdoctoral program.

\software{CLASS \citep{pety05}, GBTIDL \citep{marganian06}, NAUTILUS \citep{ruaud16}}

\newpage
\newpage

\bibliographystyle{aasjournal}
\bibliography{library_obs}

\end{document}